\documentclass[10pt]{article}
\usepackage{amssymb}
\usepackage{amsmath}
\usepackage{amsthm}
\usepackage{latexsym}
\usepackage[dvips]{epsfig}
\usepackage{mathrsfs}
\usepackage{eufrak}
\usepackage{tikz}

\theoremstyle{plain}
\newtheorem{proposition}{Proposition}
\newtheorem{lemma}{Lemma}
\newtheorem{theorem}{Theorem}

\newtheorem{corollary}{Corollary}
\newtheorem*{main}{Theorem}

\font\SYM=msbm10
\newcommand{\Real}{\mbox{\SYM R}}
\newcommand{\Complex}{\mbox{\SYM C}}
\newcommand{\Natural}{\mbox{\SYM N}}

\newcommand{\Sphere}{\mbox{\SYM S}}

\font\tenscr=rsfs10 scaled1100
\font\sevenscr=rsfs7 
\font\fivescr=rsfs5 
\skewchar\tenscr='177
\skewchar\sevenscr='177
\skewchar\fivescr='177
\newfam\scrfam
\textfont\scrfam=\tenscr
\scriptfont\scrfam=\sevenscr
\scriptscriptfont\scrfam=\fivescr

\def\scri{{\fam\scrfam I}}

\newcommand{\updn}[3]{#1^{#2}_{\phantom{#2}#3}}
\newcommand{\dnup}[3]{#1_{#2}^{\phantom{#2}#3}}

\newcommand{\dnupdn}[4]{#1_{#2 \phantom{#3} #4}^{\phantom{#2}#3}}

\def\o{o}
\def\i{\iota}
\def\Ad{{A'}}

\def\es{{\bar{s}}}
\def\er{{\bar{r}}}

\newcounter{mnote}

\begin{document}

\bibliographystyle{/home/network/jav/tex/reporthack}

\title{\textbf{A stability result for purely radiative spacetimes}}

\author{
{\Large Christian L\"ubbe} \thanks{E-mail address:
 {\tt c.luebbe@qmul.ac.uk}} \\ 
{\Large Juan Antonio Valiente Kroon} \thanks{E-mail address:
 {\tt j.a.valiente-kroon@qmul.ac.uk}} \\
School of Mathematical Sciences, Queen Mary, University of London,\\
Mile End Road, London E1 4NS, United Kingdom.}

\maketitle

\begin{abstract}
  An existence and stability result for a class of purely radiative
vacuum spacetimes arising from hyperboloidal data is given. This
result generalises semiglobal existence results for
Minkowski-like spacetimes to the case where
the reference solution contains gravitational radiation. The analysis
makes use of the extended conformal field equations and a gauge based
on conformal geodesics so that the location and structure of the
conformal boundary of the perturbed solutions is known \emph{a
priori}.
\end{abstract}


\section{Introduction}
\label{Introduction}

In this article we analyse the stability of the hyperboloidal initial
value problem for a class of solutions to the vacuum Einstein field
equations ---the so-called \emph{purely radiative vacuum
spacetimes}. Purely radiative vacuum spacetimes describe
gravitational radiation which comes from infinity, interacts with
itself in a non-linear way and then disperses to infinity. In order to
encode that the spacetime is made only of incoming gravitational
radiation, one assumes that it admits a Penrose conformal
extension such that the null generators of future and past null
infinity ($\mathscr{I}^+$ and $\mathscr{I}^-$) are complete and that
future and past timelike infinity are represented in the conformal
extension by two points $i^+$ and $i^-$. The points $i^\pm$ are
required to be regular ---that is, the conformally rescaled unphysical
spacetime admits a smooth extension which is regular at $i^\pm$. More
precisely, given a vacuum spacetime
$(\tilde{\mathcal{M}},\tilde{g}_{\mu\nu})$ it will be assumed that
there exists a conformally related spacetime
$(\mathcal{M},g_{\mu\nu})$ such that
\begin{equation}
\label{rescaled:metric}
g_{\mu\nu} =\Theta^2 \tilde{g}_{\mu\nu},
\end{equation}
where $\Theta$ is a suitable conformal factor. Null infinity,
$\mathscr{I}^\pm$, is characterised as the set of points for which
\begin{equation}
\label{scri:condition}
\Theta=0, \quad \mbox{d}\Theta  \ne 0.
\end{equation}
On the other hand, the points $i^\pm$ in a purely radiative spacetime
are such that
\begin{equation}
\label{i+:condition}
\Theta=0, \quad \mbox{d} \Theta=0, \quad \mbox{Hessian}\, \Theta \quad \mbox{is non-degenerate}.
\end{equation}
It is perhaps worth noting that the corresponding points $i^\pm$ in a
spacetime with a black hole (say, the Schwarzschild spacetime) do not
satisfy these conditions.

\medskip
It has been know for a long time that the question of the existence of
non-trivial examples of purely radiative solutions to the Einstein field equations hinges very
delicately on the structure of the spacetimes in a neighbourhood of
the so-called spatial infinity, $i^0$ ---see e.g.
\cite{Fri88,Fri92,Fri98a,Fri99}. A way of getting around the so-called
$i^0$-\emph{problem} is to consider initial value problems posed on
\emph{hyperboloidal hypersurfaces}, that is, spacelike hypersurfaces
which can be thought of as intersecting null infinity in a transversal
manner. If the initial data prescribed on the hyperboloid is suitably
close to a \emph{canonical} hyperboloidal data implied by the
Minkowski spacetime ---that is, the data is a \emph{perturbation} of
Minkowski data--- then it has been proved that the resulting
development has a conformal structure similar to that of Minkowski
spacetime \cite{Fri86b}. More precisely, the future development of
initial data admits a conformal compactification such that the locus
of points, $\mathscr{I}^+$ satisfying the condition
(\ref{scri:condition}) has the topology $\Sphere^2 \times [0,1)$, and that 
there is a single point, $i^+$, satisfying the conditions
(\ref{i+:condition}) where the null generators of null infinity
converge. The original proof of this result was carried out using the
so-called \emph{conformal Einstein field equations} discussed in
\cite{Fri81a,Fri81b,Fri82,Fri83}. These equations have
the property of being regular even at the points where the conformal
factor vanishes and are such that a solution to them implies
a solution to the vacuum Einstein field equations. A property of the
evolution systems implied by these conformal field equations is that
it includes the conformal factor as one of the unknowns, and thus any
discussion of the structure of the conformal boundary has to be
performed \emph{a posteriori}. The results of \cite{Fri86b} have been
generalised to the case of the Einstein field equations are coupled to
Maxwell and Yang-Mills fields \cite{Fri91}.

\medskip
A more general system of conformal field equations has been introduced
in \cite{Fri95}. This \emph{extended conformal Einstein field
equations} are expressed in terms of so-called \emph{Weyl connections}
---torsion free connections (not necessarily Levi-Civita) which
preserve the conformal structure. Thus, the field equations contain more gauge freedom. In particular, using the extended
conformal equations it is possible to introduce gauges ---\emph{conformal
Gaussian gauges}--- which are based on certain type of conformal
invariants ---the so-called \emph{conformal geodesics}. The advantages of
these conformal Gaussian systems are twofold: combined with the
extended conformal field equations, they imply simpler systems of
evolution equations; and, if working in vacuum, they provide an \emph{a
priori} canonical conformal factor so that the location and properties
of the conformal boundary are known and can be adjusted according to
need before the propagation equations are solved. In \cite{LueVal09},
the extended conformal Einstein field equations and conformal Gaussian
gauge systems have been used to provide a new version of the proof of
the stability results for Minkowski and de Sitter spacetime. In
particular, the analysis of the properties of the conformal boundary
of the spacetimes obtained is much simpler and transparent.

\medskip
The proofs of the stability of perturbations of the Minkowski
spacetime in \cite{Fri86b} and in \cite{LueVal09} make use of an
adaptation of a very general result on the existence and stability of
symmetric hyperbolic systems by Kato \cite{Kat70,Kat73,Kat75}. The
possibility of using these theorems to directly prove the global
existence of solutions to the Einstein field equations stems from the
use of a conformally compactified picture where timelike infinity is
at a finite position. As a result, Kato's theorems allow to provide
very compact global existence results which require almost no PDE
(partial differential equations) analysis. The latter should be
contrasted, for example, with the analysis of the non-linear stability
of the Minkowski spacetime carried out by Christodoulou \& Klainerman
\cite{ChrKla90} where precise asymptotic estimates of the solution are
obtained.

\medskip
In the present article we generalise the methods and result of
\cite{LueVal09} to reference solutions which are purely radiative
spacetimes in the sense discussed in the first paragraphs of this
introduction. A procedure to construct an infinite number of these
purely radiative spacetimes has been given in \cite{Fri88}. The
possibility of generalising the existence and stability
results of the Minkowski spacetime using this class of solutions has
been suggested in \cite{Fri91}. This construction depends upon a
remarkable connection between stationary solutions to the Einstein
field equations and spacetimes with vanishing mass. The reference
spacetimes thus obtained are only known in an abstract sense and 
without much information about their geometry at hand. It is
remarkable that although one has very little explicit information
about the reference solutions, it is still possible to use the methods
of \cite{LueVal09} to obtain a stability result. Our main result can
be summarised in the following form.

\begin{main}
Given a sequence of multipole moments for a static solution to the
Einstein vacuum equations subject to a suitable convergence condition,
there exists a hyperboloidal initial data set whose future development
is a purely radiative spacetime admitting a conformal compactification
with a future null infinity satisfying conditions
(\ref{scri:condition}) and a point representing timelike infinity
satisfying conditions (\ref{i+:condition}). A small enough
perturbation of the hyperboloidal initial data has a future development
whose conformal boundary has the same properties as that of the
unperturbed solution.
\end{main}

In order to prove this theorem one starts by considering the asymptotic
end of an initial data set for a vacuum static solution to the
Einstein field equations. This asymptotic end can be compactified to a
3-dimensional manifold, $\bar{\mathcal{S}}$, such that the point at
infinity corresponds to a certain point $i$. The
subsequent analysis will be restricted to a suitable neighbourhood of
this point. The observations in \cite{Fri88} allow to construct an
initial data set for the conformal Einstein field equations which is
analytical in a neighbourhood of $i$. This initial data has the
peculiarity of having a vanishing ADM mass. As our analysis is local
to $i$ the later feature causes no problem. A local existence for this
data follows directly. Our interest is then concentrated to the region
in the development which is included in $I^-(i)$ (the causal past of
$i$). As a consequence of our conformal setting, this region implies
the existence a portion of a purely radiative spacetime. The latter
can be regarded as the future development of some data for a purely
radiative spacetime prescribed on a fiduciary hyperboloid
$\mathring{\mathcal{H}}$. The point $i$ in $\bar{\mathcal{S}}$
corresponds to the point $i^+$ of the radiative spacetime. In order to
prove the existence of the reference radiative spacetime one employs
a conformal Gaussian system based on a congruence of conformal
geodesics leaving the hypersurface $\bar{\mathcal{S}}$
orthogonally. 

It should be noted that this congruence will, in general, intersect the
fiduciary hyperboloid in a non-orthogonal manner. This feature hinders
a direct application of Kato's theorems in order to conclude the
stability of the hyperboloidal data. The reason for this is that in
order to make use of Kato's theorems one has to make use of a Gaussian
system in which the conformal geodesics leave the initial hyperboloid
with the same angle and orientation as the congruence in the reference
solution meets the hyperboloid $\mathring{\mathcal{H}}$. It turns out
that the data for such an evolution are not really intrinsic to the
initial hyperboloid but contain information along the direction normal
to the initial hypersurface. Thus, one needs to show, first, local
existence for the perturbed hyperboloidal data ---using, for example,
a congruence of conformal geodesics which is orthogonal to the
hyperboloid. Once this local existence has been established, one can
construct initial data for the evolution along the \emph{tilted}
congruence ---using a Lorentz transformation. From here Kato's
theorems and the requirement of the perturbed data being close enough
to the reference solution yields the required stability result.

\subsection*{Outline of the article}

We start by setting our conventions in section \ref{conventions} and
recalling the physical and the conformal vacuum constraints.  As we
are working on $\Sphere^3 $ and subsets thereof, we fix coordinates
and vector fields on $\Sphere^3 $ and briefly review the norm and the
extension operator for Sobolev spaces associated to $\Sphere^3 $.  In
section \ref{CEFE} we discuss the conformal tools, which we will be
using later on. These include conformal geodesic, their associated
conformal Gaussian gauge and the conformal Einstein field
equations. We follow the standard approach by Friedrich and derive an
evolution system form the field equation by using space spinors.
Section \ref{section:reference:radiative:spacetime} is concerned with
the construction of a purely radiative reference spacetime from time
symmetric static vacuum data with non-vanishing mass given on a Cauchy
surface passing through the point $i$ at spacelike infinity.  In
section \ref{section:hyperboloidal:reference:data} the resulting
radiative spacetime is analysed from the viewpoint of hyperboloidal
data on a surface not passing through $i$. We note at this stage that
the initial data on such a hypersurface is not proper hyperboloidal
data due to the fact that the conformal geodesic congruence does not
intersect the hyperboloid orthogonally. Thus we analyse the
relationship between this so named tilted hyperboloidal data and the
proper hyperboloidal data.  Section
\ref{section:perturbed:radiative:spacetime} is concerned with the
existence and stability result for radiative spacetimes. The structure
of the conformal boundary is analysed prior to the evolution of the
data.  In section \ref{section:stability:result} we use a modified
version of a theorem by T.Kato employed in \cite{Fri86b} and
\cite{LueVal09} to establish the existence and stability for purely
radiative spacetimes in theorems \ref{Katopart2} and \ref{Katopart2a}.

\section{Basics and conventions}
\label{conventions}

We shall consider spacetimes
$(\tilde{\mathcal{M}},\tilde{g}_{\mu\nu})$ satisfying the vacuum
Einstein field equations
\begin{equation}
\tilde{R}_{\mu\nu}=0, \label{EFE}
\end{equation}
where $\tilde{R}_{\mu\nu}$ denotes the Ricci tensor of a Lorentzian
metric $\tilde{g}_{\mu\nu}$, ($\mu,\nu=0,1,2,3$) with signature
$(+,-,-,-)$.  The discussion of the solutions to equation (\ref{EFE})
will be carried out in terms of a conformally rescaled metric
$g_{\mu\nu}$ related to $\tilde{g}_{\mu\nu}$ according to equation
(\ref{rescaled:metric}). In what follows, let $\nabla_\mu$ denote the
Levi-Civita covariant derivative of $g_{\mu\nu}$, while
$R_{\mu\nu\lambda\rho}$, $C_{\mu\nu\lambda\rho}$ , $R_{\mu\nu}$, $R$
denote the associated Riemann, Weyl and Ricci tensors and the Ricci
scalar of $g_{\mu\nu}$.

In the stability analysis we will compare a spacetime $(\mathcal{M},
g_{\mu\nu})$, evolved from some chosen initial data, with a given
reference spacetime. Throughout the article, the latter will be denote
by $(\mathring{\mathcal{M}}, \mathring{g}_{\mu\nu}) $. We
will also adapt this notation for all quantities related to the
reference spacetime.

We will make use of two types of hyperboloidal data. \emph{Proper}
hyperboloidal data is given with respect to a frame consisting of
three tangent vectors to the hypersurface and its normal, while
\emph{tilted} hyperboloidal data depends on a frame not intrinsic in
the hypersurface. In our setup this frame will be oriented along the
congruence. We will distinguish between the two frames and the
corresponding frame components, by using an underscore for proper
hyperboloidal data.

The conformal factor on the spacetime will be denoted by $\Theta$, while 
$\Omega$ is the conformal factor on an initial hypersurface.

\subsection{The  constraint equations}
Let $\tilde{\mathcal{S}}$ denote a timelike hypersurface of the
spacetime $(\tilde{\mathcal{M}},\tilde{g}_{\mu\nu})$. Let
$\tilde{h}_{\alpha\beta}$, $\tilde{K}_{\alpha\beta}$ denote,
respectively, the first and second fundamental forms induced by the
metric $\tilde{g}_{\mu\nu}$ on $\tilde{\mathcal{S}}$
($\alpha,\beta=1,2,3$). The vacuum Einstein field equations (\ref{EFE})
imply the following constraint equations on $\tilde{\mathcal{S}}$
\begin{subequations}
\begin{eqnarray}
&& \tilde{r}- \tilde{K}^2 + \tilde{K}^{\alpha\beta}\tilde{K}_{\alpha\beta}=0, \label{Hamiltonian:constraint}\\
&& \tilde{D}^\alpha \tilde{K}_{\alpha\beta} -\tilde{D}_\beta \tilde{K}=0,
\label{Momentum:constraint}
\end{eqnarray}
\end{subequations}
where $\tilde{D}$ denotes the Levi-Civita and $\tilde{r}$ the
Ricci scalar of the metric $\tilde{h}_{\alpha\beta}$. The first
and second fundamental forms determined by the metrics $g_{\mu\nu}$
and $\tilde{g}_{\mu\nu}$ on $\tilde{\mathcal{S}}$ are related by
\[
h_{\alpha\beta} = \Omega^2 \tilde{h}_{\alpha\beta}, \quad K_{\alpha\beta} = \Omega \left( \tilde{K}_{\alpha\beta} + \Sigma \tilde{h}_{\alpha\beta}  \right),
\]
where $\Omega \equiv \Theta \vert_{\tilde{\mathcal{S}}} $ and the
function $\Sigma$ on $\tilde{\mathcal{S}}$ denotes the derivative of
$\Theta$ in the direction of the future directed $g$-normal of
$\tilde{\mathcal{S}}$. The traces
$K=h^{\alpha\beta}K_{\alpha\beta}$ and
$\tilde{K}=\tilde{h}^{\alpha\beta}\tilde{K}_{\alpha\beta}$ are
related by
\[
\Omega K = \tilde{K} + 3\Sigma.
\]
The Hamiltonian and momentum constraints for vacuum, equations
(\ref{Hamiltonian:constraint}) and (\ref{Momentum:constraint}),
written in terms of the conformal fields $\Omega$, $\Sigma$,
$h_{\alpha\beta}$ and $K_{\alpha\beta}$ read
\begin{subequations}
\begin{eqnarray}
&& 2\Omega D_\alpha D^\alpha \Omega -3 D_\alpha \Omega D^\alpha \Omega +\frac{1}{2} \Omega^2 r - 3\Sigma^2 - \frac{1}{2}\Omega^2 \left( K^2 -K_{\alpha\beta} K^{\alpha\beta} \right) + 2\Omega \Sigma K =0, \label{conformal_Hamiltonian}\\
&& \Omega^3 D^\alpha \left(\Omega^{-2} K_{\alpha\beta}\right) -\Omega \left( D_\beta K -2\Omega^{-1} D_\beta \Sigma \right)=0, \label{conformal_Momentum}
\end{eqnarray}
\end{subequations}
where $D$ denotes the Levi-Civita connection and $r$ the Ricci scalar of
the metric $h_{\alpha\beta}$. In particular if $\Omega=0$ one has that
\begin{equation}
\label{cH:boundary}
\Sigma^2+ D_\alpha \Omega D^\alpha \Omega=0.
\end{equation}

\bigskip In the sequel we will consider two different classes of
solutions to equations (\ref{conformal_Hamiltonian}) and
(\ref{conformal_Momentum}). The first class will consist of initial
data sets for the Einstein vacuum field equations which are
\emph{asymptotically Euclidean}, and thus, the relevant initial
hypersurfaces are Cauchy hypersurfaces of an asymptotically flat
spacetime. Further details will be discussed in section
\ref{Asympt:Euclidean:regular}. The second class of solutions to be
considered are hyperboloidal data sets. In this case the initial
hypersurfaces are not Cauchy hypersurfaces of a globally hyperbolic
spacetimes. Further details will be given in section
\ref{section:hyperboloidal:data}. The character of the solutions to
the conformal constraint equations (asymptotically Euclidean or
hyperboloidal) is determined through the boundary conditions which
supplement equations (\ref{conformal_Hamiltonian}) and
(\ref{conformal_Momentum}).

\subsection{Coordinates on $\Sphere^3$ and submanifolds thereof}
\label{coordinates:S3}

In the sequel we shall work with conformal spacetimes with time
slices diffeomorphic to simply connected subsets of the 3-sphere,
$\Sphere^3$. Therefore, it will be important to have a good
coordinatisation of $\Sphere^3$ and to have a frame which is globally
defined over this manifold. Following the work in
\cite{Fri88,LueVal09} we regard $\Sphere^3$ as a submanifold of
$\Real^4$:
\[
\Sphere^3 =\left \{ x^\mathcal{A} \in \Real^4 \;\bigg | \; (x^1)^2 + (x^2)^2 + (x^3)^2 + (x^4)^4=1  \right\}.
\]
The restrictions of the functions $x^\mathcal{A}$,
$\mathcal{A}=1,2,3,4$ on $\Real^4$ to $\Sphere^3$ will again be
denoted by $x^\mathcal{A}$. If we view the point $N \equiv (0,0,0,1)$
as the North pole and $\mathcal{D} \equiv \{x^4 > 0 \} \cap \Sphere^3
$ as the open upper half sphere, then for any subset $\mathcal{U}$ of
$\mathcal{D} $ with $N \in \mathcal{U} $ the functions $(x^1, x^2, x^3) $ provide coordinates on
$\mathcal{U}$ centered at $N$. The vector fields
\begin{subequations}
\begin{eqnarray}
&& c_1 \equiv x^1 \partial_4 -x^4\partial_1 +x^2\partial_3 -x^3\partial_2, \label{v_field_c1} \\
&& c_2 \equiv x^1 \partial_3 -x^3\partial_1 +x^4\partial_2 -x^2\partial_4, \label{v_field_c2}\\
&& c_3 \equiv x^1 \partial_2 -x^2\partial_1 +x^3\partial_4 -x^4\partial_3. \label{v_field_c3}
\end{eqnarray}
\end{subequations}
on $\Real^4$ are tangent to $\Sphere^3$ and provide a globally defined
frame field on $\Sphere^3$.

\medskip
Given a simply connected 3-manifold $\mathcal{H}$ with boundary
$\partial\mathcal{H}$, there exists a diffeomorphism $\phi: \mathcal{H}
\rightarrow \mathcal{U} \equiv \{ x^4 \ge 1/2 \} \subset
\mathcal{D} $, such that $\phi: \partial\mathcal{H} \rightarrow \{ x^4 =
1/2 \} $. Since $\phi$ is a diffeomorphism we can map any
tensor from $\mathcal{H}$ to $\mathcal{U}$ and vice versa.  For
convenience we denote tensors on $\mathcal{H}$ and $\mathcal{U}$ by
the same symbols and the underlying manifold is to be understood from
the context. In particular we can pull-back the functions
$x^\mathcal{A}$ to provide coordinates on $\mathcal{H}$ and use the
vectors $c_1,\;c_2,\;c_3$, as given by
(\ref{v_field_c1})-(\ref{v_field_c3}), as a globally defined frame
field on $\mathcal{H}$.

\begin{figure}[t]
\centerline{\includegraphics[width=.7\textwidth]{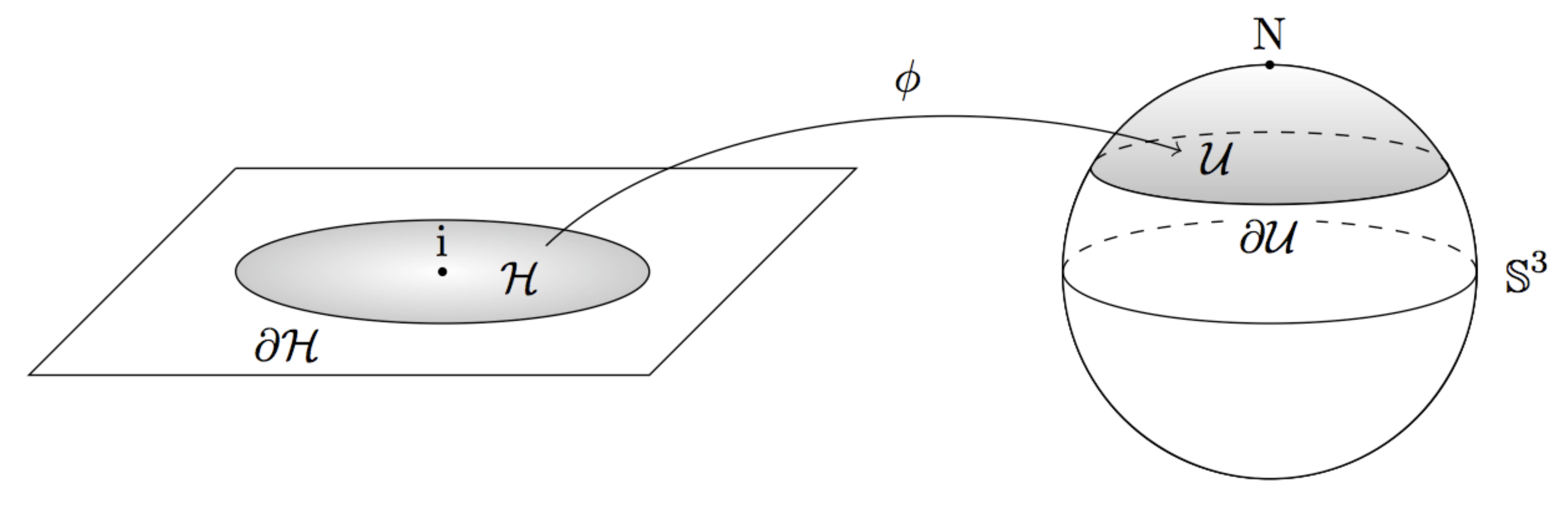}}
\caption{The mapping $\mathcal{H}$ onto $\Sphere^3$.}
\end{figure}

\subsection{Extending fields to $\Sphere^3$}
\label{section extending initial data}

In the sequel it will be necessary to extend certain (vector valued)
functions defined on a subset $\mathcal{U}\subset \Sphere^3$ to the
whole of $\Sphere^3$. For this we follow the approach pursued in
\cite{Fri86b}. On the spaces $C^\infty(\mathcal{U},\Real^N)$ and 
$C^\infty(\Sphere^3,\Real^N)$ of smooth $\Real^N$-valued functions on  $\mathcal{U}$ and $\Sphere^3$ define for $m\in \Natural$
the Sobolev-like norm
\begin{equation}
\label{norm}
\parallel w \parallel_{m,\mathcal{Q}}=\left( \sum_{k=0}^m \int_{\mathcal{Q}} \left( \sum^3_{\er_1,\ldots, \er_k=1} |D_{\er_1}\cdots D_{\er_k}w|^2 \right) \mbox{d}\mu \right)^{1/2},
\end{equation}
with $\mathcal{Q}=\mathcal{U}, \;\Sphere^3$. Let
$H^m(\mathcal{U},\Real^N)$ and $H^m(\Sphere^3,\Real^N)$ be the Hilbert
spaces obtained, respectively, as the completion of the spaces
$C^\infty(\mathcal{U},\Real^N)$ and $C^\infty(\Sphere^3,\Real^N)$ on
the norms given by formula (\ref{norm}). Given $w\in
H^m(\mathcal{U},\Real^N)$, there exists a linear extension operator
$E$ such that 
\begin{subequations}
\begin{eqnarray}
&&(Ew)(x) = w(x) \quad \mbox{ almost everywhere in } \mathcal{U} \\
\label{norm:relationship}
&&\parallel E w \parallel_{m, \Sphere^3} \leq  K \parallel w \parallel_{m,\mathcal{U}}, 
\end{eqnarray}
\end{subequations}
with $K$ a constant which is universal for fixed $m$. It is important
to notice that if, for example, $w=(h_{ij},\chi_{ij},\Omega,\Sigma)$
is a solution to the conformal constraints, equations
(\ref{conformal_Hamiltonian}) and (\ref{conformal_Momentum}), on
$\mathcal{U}$ then $Ew$ will not satisfy the constraints in
$\Sphere^3\setminus \mathcal{U}$.


\section{The conformal Einstein field equations  and conformal Gaussian gauges}
\label{CEFE}

When discussing the existence of solutions to the Einstein field
equations in terms of conformally rescaled spacetimes, it is
convenient to introduce gauges based on conformally invariant
structures. The particular conformal structures to be used in our
analysis are \emph{conformal geodesics}. These curves are autoparallel
with respect to \emph{Weyl connections} ---that is, torsion-free
connections which preserve the conformal structure, but are not
necessarily the Levi-Civita connection of a metric. Conformal geodesics
were first applied in the context of General Relativity in
\cite{FriSch87}. The idea of conformal Gaussian
coordinates based on congruences of conformal geodesics has been
introduced in \cite{Fri95,Fri98a}. It has been shown in \cite{Fri03c} that conformal
Gaussian coordinates can be used to cover even the strong field
regions of the Schwarzschild spacetime. 

\medskip 
In this section we present a brief discussion of the properties of
gauges based on conformal Gaussian coordinates, and of the associated
propagation equations implied by the extended conformal Einstein field
equations. For further details we refer the reader to
\cite{Fri03c,LueVal09,TodLue07} and to the references mentioned in the
previous paragraph.

\subsection{Conformal geodesics}

Given a spacetime $(\tilde{\mathcal{M}},\tilde{g}_{\mu\nu})$, a
conformal geodesic $x^\mu(\tau)$ and its associated 1-form
$b_\mu(\tau) $ are solutions to the system of equations
\begin{subequations}
\begin{eqnarray}
\label{cge1}
&& \dot{x}^\nu \tilde{\nabla}_\nu \dot{x}^\mu + 2 \langle b, \dot{x} \rangle \dot{x}^\mu - \tilde{g}(\dot{x}, \dot{x}) b^\mu =0,\\
\label{cge2}
&& \dot{x}^\nu \tilde{\nabla}_\nu b_\mu - \langle b, \dot{x} \rangle b_\mu + \frac{1}{2} \tilde{g}^\sharp(b,b) \dot{x}_\mu = \tilde{L}_{\lambda\mu} \dot{x}^\lambda,
\end{eqnarray}
\end{subequations}
where $\tilde{L}_{\mu\nu}$ denotes the \emph{Schouten tensor} of the
Levi-Civita connection $\tilde{\nabla}$, given by
\[
\tilde{L}_{\mu\nu}\equiv \frac{1}{2}\tilde{R}_{\mu\nu} - \frac{1}{12}\tilde{R} g_{\mu\nu}.
\]
The notation
\[
\langle b, \dot{x} \rangle \equiv b_\mu \dot{x}^\mu, \quad  \tilde{g}^\sharp(b,b) \equiv \tilde{g}^{\mu\nu} b_\mu b_\nu
\]
has been used. Associated to the pair $(x^\mu(\tau),b_\mu(\tau))$, there is a
preferred family of conformal factors, $\Theta$, which are developed
along the congruence using
\begin{equation}
\label{conformal factor}
\dot{x}^\mu \tilde{\nabla}_{\mu} \Theta = \Theta \langle b, \dot{x} \rangle, \quad \Theta|_{\tilde{\mathcal{S}}}=\Theta_*.
\end{equation}
Consistent with equation (\ref{rescaled:metric}), define
$g_{\mu\nu}\equiv \Theta^2 \tilde{g}_{\mu\nu}$, and denote the
corresponding Levi-Civita connection by $\nabla$. In this conformal gauge the
1-form along the conformal geodesics is given by
\[
f \equiv b-\Theta^{-1}\mathrm{d}\Theta
\]
and satisfies $\langle f, \dot{x} \rangle =0 $. In the following, we
will work with the unphysical metric $g_{\mu\nu}$ and the 1-form
\begin{equation}
\label{d:decomposition}
d_\mu\equiv \Theta f_\mu + \nabla_\mu \Theta.
\end{equation}
The 1-forms $b_\mu$ and $f_\mu$ are
used to define a Weyl connection, $\hat{\nabla}$, via 
\begin{eqnarray}
&&\dnupdn{\hat{\Gamma}}{\mu}{\nu}{\rho} \equiv \dnupdn{\tilde{\Gamma}}{\mu}{\nu}{\rho} + \left( \updn{\delta}{\nu}{\mu}b_\rho + \updn{\delta}{\nu}{\rho}b_\mu -g_{\mu\rho}g^{\nu\lambda}b_\lambda \right) \nonumber\\
\label{Weyl:connection}
&& \phantom{\dnupdn{\hat{\Gamma}}{\mu}{\nu}{\rho}} \equiv \dnupdn{\Gamma}{\mu}{\nu}{\rho} + \left( \updn{\delta}{\nu}{\mu}f_\rho + \updn{\delta}{\nu}{\rho}f_\mu -g_{\mu\rho}g^{\nu\lambda}f_\lambda \right).
\end{eqnarray}
In the gauge corresponding to the connection $\hat{\nabla}$ the
equations (\ref{cge1}) and (\ref{cge2}) take the simpler form
\[
\dot{x}^\nu \hat{\nabla}_\nu \dot{x}^\mu = 0, \quad \hat{L}_{\lambda\mu} \dot{x}^\lambda = 0,
\]
where $\hat{L}_{\mu\nu}$ is the Schouten tensor of the Weyl connection
$\hat{\nabla}$. It is given by
\[
\hat{L}_{\mu\nu} = L_{\mu\nu}  -\nabla_\mu f_\nu + f_\mu f_\nu  - \frac{1}{2} g_{\mu\nu} f_\lambda f^\lambda.
\]
Note that generally one will have that $\hat{L}_{[\mu\nu]}\neq 0$.

\bigskip 
We transport an orthogonal frame $e_k$, $k=0,1,2,3$, such that $e_0= \dot{x}
$, along the congruence of conformal geodesics according to
\begin{equation}
\label{propagation:frame}
\dot{x}^\mu \hat{\nabla}_\mu e_k =0,
\end{equation}
which implies that the frame is automatically $g$-orthonormal. If $\tilde{g}$ is a vacuum metric, then the conformal factor, $\Theta$, can be written down explicitly by
\begin{equation}
\label{general:Theta}
\Theta = \Theta_* + \dot{\Theta}_* (\tau-\tau_*) + \frac{1}{2} \ddot{\Theta}_*(\tau-\tau_*)^2,
\end{equation}
where $\Theta_*$, $\dot{\Theta}_*$ and $\ddot{\Theta}_*$ are functions which are constant along a given conformal geodesic. They are subject to the constraint
\begin{equation}
\label{Theta:constraint}
2\ddot{\Theta}_* \Theta_* = g^\sharp(d_*,d_*).
\end{equation}
Furthermore, along each conformal geodesic
\[
d_a = \langle d, e_a \rangle, 
\]
 are constant for $a=1,2,3$ ---see e.g. \cite{Fri03c}.

\bigskip
Analogous to the case of metric geodesic congruences there exist
deviation vector fields for congruences of conformal geodesics
\cite{Fri03c}, referred to as \emph{conformal Jacobi fields} and denote
them by $\eta^\mu$. Let $\eta_k\equiv \eta_\mu e^\mu_k$. These fields
satisfy
\begin{equation}
\label{Jacobiequs}
\hat{\nabla}_{\dot{x}} \eta_k = g(\hat{\nabla}_\eta \dot{x}, e_k).
\end{equation}
We say the congruence has a conjugate point if $\eta_a \, (a=1,2,3)$ vanish at that point. 

\subsection{Conformal Gaussian gauges and space spinors}
\label{section:conformal:gaussian:gauges}
The coordinates $x^{\mathcal{A}}$ described in section \ref{coordinates:S3}
will be extended off a fiduciary hypersurface by dragging them along a
congruence of conformal geodesics. Hence, setting $x^0=\tau$ one
obtains a coordinatisation of a slab of the fiduciary hypersurface. In
these coordinates one has that
\begin{equation}
\label{Gauss:gauge1}
e_0=\dot{x}=\partial_\tau.
\end{equation}
Let $\dnupdn{\hat{\Gamma}}{i}{j}{k}$ and $\hat{L}_{ij}$ denote,
respectively, the components of
$\dnupdn{\hat{\Gamma}}{\mu}{\nu}{\rho}$ and $\hat{L}_{\mu\nu}$ with
respect to the frame $e_k$ propagated according to the rule
(\ref{propagation:frame}). One then has that
\begin{equation}
\label{Gauss:gauge2}
\dnupdn{\hat{\Gamma}}{0}{j}{k}=0, \quad \hat{L}_{0k}=0,
\end{equation}
---see e.g. \cite{Fri95,Fri98a}. This choice of coordinates, frame
field and conformal will be referred to as a \emph{conformal Gaussian
gauge system} based on the fiduciary hypersurface. The appearance of
conjugate points in the congruence of conformal geodesics implies a
breakdown of the conformal Gaussian gauge. 

\bigskip
 Let $\tau^\mu\equiv \sqrt{2} \dot{x}^\mu$, and let $\tau^{AA'}$ be
its spinorial counterpart. In terms of spinors, the gauge conditions
(\ref{Gauss:gauge1}) and (\ref{Gauss:gauge2}) can be written as ---see
\cite{Fri04} for details of the definition of $\Theta_{AA'BB'} $---
\[
\tau^{AA'}e_{AA'} =\sqrt{2}\partial_\tau, \quad \tau^{AA'} \dnupdn{\hat{\Gamma}}{AA'}{B}{C}=0, \quad \tau^{BB'} \Theta_{AA'BB'}=0.
\]
The frame vectors $e_{AA'}$ will be expressed with respect to the
vectors $\{c_0,c_1,c_2,c_3\}$ where $c_0$ is the normal to the
surfaces of constant $\tau$ 
and $c_1$, $c_2$ and $c_3$ are given by
(\ref{v_field_c1})-(\ref{v_field_c3}). We will write
\[
e_{AA'}= e_{AA'}^\es c_\es, \quad \es=0,1,2,3.
\]

\bigskip
In order to obtain a suitable set of evolution equations we consider a
space-spinor formalism ---see e.g. \cite{Som80,Fri04,LueVal09} ---
as it allows us to obtain evolution equations involving only \emph{unprimed} spinors. 
Introduce a spin dyad $\{\o, \iota\}$ such that
\[
\tau^{A\Ad} = \o^A \bar{\o}^\Ad +  \i^A \bar{\i}^\Ad.
\]
We decompose the connection as follows
\begin{subequations}
\begin{eqnarray} 
\label{connection:ss:decomposition:1} 
&& \nabla_{AA'} = \frac{1}{2}\tau_{AA'}\nabla  - \updn{\tau}{C}{A'}\nabla_{AC},\\
\label{connection:ss:decomposition:2}
&& \nabla_{AB}\equiv \dnup{\tau}{(A}{B'} \nabla_{B)B'}, \quad \quad \nabla \equiv  \tau^{CC'}\nabla_{CC'}. 
\end{eqnarray}
\end{subequations}
Note that $\nabla_{AB}$ is not the Levi-Civita connection of the surfaces of
constant $\tau$, but the so-called \emph{Sen connection}.
The frame fields $e^\es_{AA'}$ are similarly decomposed using
\begin{subequations}
\begin{eqnarray}
&& e^\es_{AA'} = \frac{1}{2}e^\es_{CC'} \tau^{CC'}\tau_{AA'} - \updn{\tau}{C}{A'}e^\es_{AC}, \label{space:spinor:decomposition:1} \\
&& e^\es_{AB}\equiv \dnup{\tau}{(A}{B'} e^\es_{B)B'}. \label{space:spinor:decomposition:2}
\end{eqnarray}
\end{subequations}
The fields $e^\es_{AB}$ are associated to spatial vectors, and hence,
they satisfy the reality conditions
\begin{equation}
e^\es_{AB} = -\dnup{\tau}{A}{A'}
\dnup{\tau}{B}{B'}\overline{e}^\es_{A'B'}. \label{reality:condition}
\end{equation}
Contracting with $\tau^{A\Ad}$ and symmetrising, one decomposes the remaining connection and curvature spinors as follows. 
The space spinor $\Theta_{ABCD}=\Theta_{AB(CD)}$ is defined by 
\[
\Theta_{ABCD} \equiv \Theta_{AA'CC'} \updn{\tau}{A'}{B} \updn{\tau}{C'}{D}=\Theta_{(AB)(CD)} + \frac{1}{2} \epsilon_{AB} \dnupdn{\Theta}{G}{G}{(CD)}.
\]
One observes that $\hat{\Gamma}_{AA'BC} = \Gamma_{AA'BC} + \epsilon_{AB}f_{CA'}$ and $\tau^{A\Ad}f_{A\Ad}=0 $.
Define $\Gamma_{ABCD} \equiv \dnup{\tau}{B}{B'} \Gamma_{AB'CD}$,
which in turn, will be decomposed as 
\[
\Gamma_{ABCD} = \frac{1}{\sqrt{2}}\left(\xi_{ABCD} - \chi_{(AB)CD} \right) -\frac{1}{2}\epsilon_{AB} f_{CD}.
\]
The spinors in the latter equation possess the following symmetries
\[
\Gamma_{ABCD}=\Gamma_{AB(CD)}, \quad \chi_{ABCD}=\chi_{AB(CD)}, \quad \xi_{ABCD}=\xi_{(AB)(CD)}.
\]

\subsection{The conformal Einstein evolution equations}
\label{conformal:evolution:equations}

In the sequel we will make use of propagation equations which are derived from the extended conformal field equations of \cite{Fri95}
---see also \cite{Fri03a,Fri04}. 
These equations are a generalisation of the original conformal equations which allow the use of Weyl
connections, as defined in (\ref{Weyl:connection}). 
Weyl connections make it possible to consider more general gauges in the derivation of systems of propagation equations out of the conformal field equations. 
In particular, it makes it possible to make use of the conformal Gaussian gauge systems discussed in section \ref{section:conformal:gaussian:gauges}. 

Using the space spinor formalism and suitable contractions of the extended conformal field equations
with $\tau^{BB'}$, one finds that the conformal Gaussian gauge system implies the
following propagation equations for the unknowns $e^\es_{AB}$,
$\xi_{ABCD}$, $f_{AB}$, $\chi_{(AB)CD}$, $\Theta_{(AB)CD}$ and 
$\Theta_{G\phantom{G}CD}^{\phantom{G}G}$:

\begin{subequations}
\begin{eqnarray}
&&\partial_\tau e^0_{AB}=-\dnup{\chi}{(AB)}{EF}e^{0}_{EF}-f_{AB}, \label{p1} \\
&&\partial_\tau e^\er_{AB}=-\dnup{\chi}{(AB)}{EF}e^\er_{EF}, \quad \er=1,2,3 \label{p2}\\
&&\partial_\tau \xi_{ABCD}=-\dnup{\chi}{(AB)}{EF}\xi_{EFCD}+\frac{1}{\sqrt{2}}(\epsilon_{AC}\chi_{(BD)EF}+\epsilon_{BD}\chi_{(AC)EF})f^{EF} \nonumber\\
&&\hspace{2cm} -\sqrt{2}\dnup{\chi}{(AB)(C}{E}f_{D)E}-\frac{1}{2}(\epsilon_{AC}\dnupdn{\Theta}{F}{F}{BD}+\epsilon_{BD}\dnupdn{\Theta}{F}{F}{AC})-\mbox{i}\Theta\mu_{ABCD}, \label{p3} \\
&&\partial_\tau f_{AB}=-\dnup{\chi}{(AB)}{EF}f_{EF}+\frac{1}{\sqrt{2}}\dnupdn{\Theta}{F}{F}{AB}, \label{p4} \\
&&\partial_\tau \chi_{(AB)CD}=-\dnup{\chi}{(AB)}{EF}\chi_{EFCD}-\Theta_{(CD)AB}+\Theta\eta_{ABCD}, \label{p5} \\
&&\partial_\tau\Theta_{(AB)CD}=-\dnup{\chi}{(CD)}{EF}\Theta_{(AB)EF}-\partial_\tau\Theta\eta_{ABCD}+\mbox{i}\sqrt{2}\updn{d}{E}{(A}\mu_{B)CDE}, \label{p6} \\
&&\partial_\tau \dnupdn{\Theta}{G}{G}{AB}=-\dnup{\chi}{(AB)}{EF}\dnupdn{\Theta}{G}{G}{EF}+\sqrt{2}d^{EF}\eta_{ABEF}, \label{p7}
\end{eqnarray}
\end{subequations}
where $\eta_{ABCD}$ and $\mu_{ABCD}$ denote, respectively, the
electric and magnetic parts of $\phi_{ABCD}$ and $d_{AB} $ is the
spinor representation of $d_{AA'}$ ---for further details see
e.g. \cite{Fri98a,Fri04,LueVal09}. Note that the above equations
involve an unspecified conformal factor $\Theta$ and the space spinor
representation of the 1-form $d_\mu$ ($d_{AB}$) introduced in
(\ref{d:decomposition}). For $\Theta$ we will use (\ref{conformal
factor}) and (\ref{general:Theta}).

\bigskip
The evolution equations for the spinor $\phi_{ABCD}$ are derived from a space-spinor decomposition of the Bianchi identity
\begin{equation}
\label{spinor:Bianchi}
\nabla^{AA'}\phi_{ABCD}=0.
\end{equation}
The resulting propagation equations are not unique as one can always
add to them a multiple of the constraint equations implied by
(\ref{spinor:Bianchi}), $\nabla^{AB}\phi_{ABCD}=0$.  In
\cite{Fri95,Fri98a} the following Bianchi propagations equations
\begin{equation}
\label{extended_cfe4}
\sqrt{2}\partial_\tau \phi_{ABCD} - 2 \dnup{\nabla}{(D}{F}\phi_{ABC)F}=0.
\end{equation}
are considered ---to the so called \emph{standard system}.
Note that equation (\ref{extended_cfe4}) is expressed with respect to the
Levi-Civita connection $\nabla$ and not the Weyl connection $\hat{\nabla}$.

\bigskip
In order to keep track of the possible appearance of conjugate points
in the solutions to the propagation equations, we append to
(\ref{p1})-(\ref{p7}) and (\ref{extended_cfe4}) an evolution equation
for the Jacobi field. The conformal Jacobi field $\eta^\mu$ has a
spinorial counterpart $\eta_{AA'}$ which can be split as
\[
\eta_{AA'} = \frac{1}{2}\eta \tau_{AA'} -\updn{\tau}{B}{A'} \eta_{AB}, \quad 
\eta \equiv \eta_{AA'} \tau^{AA'}, \quad \eta_{AB}\equiv \dnup{\tau}{(A}{B'} \eta_{B)B'}.
\]
Conjugate points in the congruence of conformal geodesics arise if
$\eta_{AB}=0$. Equation (\ref{Jacobiequs}) implies that the fields
$\eta$, $\eta_{AB}$ satisfy the propagation equations
\begin{subequations}
\begin{eqnarray}
&& \sqrt{2} \partial_\tau \eta= f_{AB} \eta^{AB}, \label{j1}\\
&& \sqrt{2} \partial_\tau \eta_{AB} = \chi_{CD(AB)} \eta^{CD}. \label{j2}
\end{eqnarray}
\end{subequations}

\bigskip
Besides the above propagation equations (\ref{p1})-(\ref{p7}), and
(\ref{extended_cfe4}) one also obtains a set of equations referred to
as the \emph{extended conformal constraint equations} involving the
operators $e_{AB}$ and $\hat{\nabla}_{AB}$ (respectively
$\nabla_{AB}$).  It should be noticed that despite their name, the
constraint equations, are not intrinsic to the hypersurfaces of
constant $\tau$ ---except, possibly, for an initial fiduciary slice---
as generically, they contain $\partial_\tau$ derivatives. The crucial
observation for our purposes concerns their \emph{propagation properties} as
given by the following lemma proved in \cite{Fri95}. Let
$D^+(\mathcal{S})$ denote the future domain of dependence of
$\mathcal{S}$ ---see e.g. \cite{HawEll73,Wal84} for definitions.

\begin{lemma}
\label{lemma:propagation constraints}
Assume that the constraint equations are satisfied on an
initial hypersurface $\mathcal{S}$. If the conformal propagation
equations are satisfied on $D^+(\mathcal{S})$, then the conformal constraint
equations are also satisfied on $D^+(\mathcal{S})$.
\end{lemma}

\subsection{Structural properties of the evolution equations}
\label{section:structural:properties}
We discuss now some general structural properties of the equations
(\ref{p1})-(\ref{p7}), (\ref{extended_cfe4}), (\ref{j1})-(\ref{j2})
which will be used systematically in the sequel. Let 
\[
\phi_i \equiv \phi_{(ABCD)_i},
\]
where the subscript ${}_{(ABCD)_i}$ indicates that after
symmetrisation $i$ indices are set to $1$. Introduce the
notation
\[
\upsilon \equiv \left(e^\es_{AB}, \Gamma_{ABCD}, \Theta_{ABCD},\eta,\eta_{AB}\right), \quad \phi\equiv \left(\phi_0,\phi_1,\phi_2,\phi_3,\phi_4\right),
\]
where it is understood that $\upsilon$ contains only the independent
components of the respective spinor ---which are obtained by writing
linear combinations of irreducible spinors. In terms of
$\upsilon$ and $\phi$, the propagation equations (\ref{p1})-(\ref{p7})
and (\ref{j1})-(\ref{j2}) can be written as:
\begin{equation}
\label{upsilon:propagation}
\partial_\tau \upsilon = K\upsilon + Q(\upsilon,\upsilon)+ L\phi,
\end{equation}
where $K$ and $Q$ denote , respectively, a linear constant
matrix-valued function and a bilinear vector-valued function both with
constant entries and $L$ is a linear matrix-valued function with
coefficients depending on the coordinates.  Similarly equation
(\ref{extended_cfe4}) can be written in the form
\begin{equation}
\sqrt{2}E \partial_\tau \phi + A^{AB} e^\es_{AB}\partial_\es \phi =B(\Gamma_{ABCD})\phi, \label{bianchi:propagation}
\end{equation}
where $E$ denotes the $5\times 5$ identity matrix and $A^{AB}e^\es_{AB}$,
$\es=0,\ldots,3$, are $5\times 5$ matrices depending on the
coordinates, while $B(\Gamma_{ABCD})$ denotes a constant matrix-valued
linear function of the connection coefficients $\Gamma_{ABCD}$. For
later reference it is noted that
\begin{equation}
\sqrt{2}E + A^{AB}e^0_{AB}=
\left(
\begin{array}{ccccc}
\sqrt{2}-2e^0_{01} & 2e^0_{00} & 0 & 0 & 0 \\
-e^0_{11} & \sqrt{2} & e^0_{00} & 0 & 0 \\
0 & -e^0_{11} & \sqrt{2} & e^0_{00} & 0 \\
0 & 0 & -e^0_{11} & \sqrt{2} & e^0_{00} \\
0 & 0 & 0 & -2e^0_{11} & \sqrt{2} + 2e^0_{01}
\end{array}
\right), \label{matrix:A0}
\end{equation}
and that
\[
A^{AB}e^\er_{AB}=
\left(
\begin{array}{ccccc}
-2e^\er_{01} & 2e^\er_{00} & 0 & 0 & 0 \\
-e^\er_{11} & \sqrt{2} & e^\er_{00} & 0 & 0 \\
0 & -e^\er_{11} & \sqrt{2} & e^\er_{00} & 0 \\
0 & 0 & -e^\er_{11} & \sqrt{2} & e^\er_{00} \\
0 & 0 & 0 & -2e^\er_{11} & 2e^\er_{01}
\end{array}
\right),
\]
with $\er=1,2,3$. From the reality condition (\ref{reality:condition})
one has that
\begin{eqnarray*}
&& e^0_{00}=-\overline{e}^0_{11}, \quad e^\er_{00}=-\overline{e}^\er_{11}, \\
&& e^0_{01}=\overline{e}^0_{01}, \quad e^\er_{01}=\overline{e}^\er_{01},
\end{eqnarray*}
so that in particular $e^0_{01}$ and $e^\er_{01}$ are real. The
Hermitian matrices $\sqrt{2}E + A^{AB}e^0_{AB}$, $A^{AB}e^\er_{AB}$
imply real symmetric matrices for the propagation equations obtained
from splitting the system
(\ref{upsilon:propagation})-(\ref{bianchi:propagation}) into real and
imaginary parts of the propagation equations. These matrices are
explicitly given in \cite{LueVal09}.

\bigskip Let
$u\equiv(\mbox{Re}(\upsilon),\mbox{Im}(\upsilon),\mbox{Re}(\phi),\mbox{Im}(\phi))$.
The unknown $u$ takes values in $\Real^N$ for some $N\in \Natural$. In this article 
$u$ will be defined over $\Sphere^3 $ or subsets thereof.
From equations (\ref{upsilon:propagation}) and
(\ref{bianchi:propagation}) it follows that $u$ satisfies a system of
quasilinear partial differential equations for $u$ of the form
\begin{equation}
\label{sh:system}
A^0(u) \cdot \partial_\tau u+ \sum^3_{\er=1} A^\er(u)\cdot c_\er(u) + B(\tau,x^{\mathcal{A}},u)\cdot u=0,
\end{equation} 
with $c_\er(u)$ denoting the vector fields
(\ref{v_field_c1})-(\ref{v_field_c3}) acting on the unknown $u$. Given
any $z\in \Real^N$, the matrix valued functions $A^\es(z)$,
$\es=0,1,2,3$ have entries which are polynomial in $z$. These
polynomials are at most of degree one and have constant coefficients.
The matrices are symmetric ${}^t(A^\es(z))=A^\es(z)$, $z\in\Real^N$.
The matrix valued function $B=B(\tau,x^\mathcal{A},z)$ with
$(\tau,x^\mathcal{A},z)\in \Real\times \Sphere^3 \times \Real^N$ has
entries which are polynomials in $z$ (of at most degree 1) with
coefficients which are analytic functions on $\Real \times
\Sphere^3$.

\section{Purely radiative reference spacetimes}
\label{section:reference:radiative:spacetime}
In what follows we will construct purely radiative vacuum
spacetimes using the ideas of \cite{Fri88}. These spacetimes are
required to allow a conformal compactification such that the resulting
conformal metric is smooth at null infinity including the point past
or future timelike infinity. The purely radiative spacetimes in
\cite{Fri88} have been obtained from of static, asymptotically flat
solutions to the Einstein field equations. Before discussing the
results in \cite{Fri88} which are crucial for our analysis, we
introduce some definitions.

\subsection{Asymptotically Euclidean hypersurfaces}
\label{Asympt:Euclidean:regular}
Asymptotically flat spacetimes,
$(\tilde{\mathcal{M}},\tilde{g}_{\mu\nu})$, can be obtained as the
development of \emph{asymptotically Euclidean} initial data sets. The
initial data
$(\tilde{\mathcal{S}},\tilde{h}_{\alpha\beta},\tilde{K}_{\alpha\beta})$
will be said to be asymptotically Euclidean if there is a compact
subset of $\tilde{\mathcal{S}}$ whose complement for some positive
number $r_0$ is diffeomorphic to $\{ y^\alpha \in \Real^3 \; | \; |y|
> r_0\}$, where $|y|^2 \equiv (y^1)^2 +(y^2)^2 +(y^3)^2$. In terms of
the coordinates $y^\alpha$ introduced by this identification a
standard asymptotic flatness requirement is
\[
\tilde{h}_{\alpha\beta} =-\left(1+\frac{2m}{|y|}  \right) \delta_{\alpha\beta} + \mathcal{O}\left( \frac{1}{|y|^2}\right), \quad \mbox{ as } |y|\rightarrow \infty,
\]
with $m$ a constant ---the ADM mass of $\tilde{\mathcal{S}}$. An
initial data set
$(\tilde{\mathcal{S}},\tilde{h}_{\alpha\beta},\tilde{K}_{\alpha\beta})$
is said to be time symmetric if $\tilde{K}_{\alpha\beta}=0$. It follows form
the Einstein field equations that its development will have a time
reflexion symmetry with respect to the surface $\tilde{\mathcal{S}}$.

\medskip
As we
shall be working with conformally compactified spacetimes, it will
convenient to consider a compactified version of the initial
hypersurface $\tilde{\mathcal{S}}$. To this end, it will be assumed
that there is a 3-dimensional, orientable, smooth \emph{compact}
manifold $(\bar{\mathcal{S}},h)$, a point $i \in \bar{\mathcal{S}}$, a
diffeomorphism $\Phi:\bar{\mathcal{S}}\setminus \{i\} \longrightarrow
\tilde{\mathcal{S}}$ and a function $\Omega \in
C^2(\bar{\mathcal{S}}) \cap C^\infty(\tilde{\mathcal{S}})$ with the
properties
\begin{subequations}
\begin{eqnarray}
&& \Omega(i)=0, \quad D_{\alpha}\Omega(i)=0, \quad D_\alpha D_\beta \Omega(i)=-2h_{\alpha\beta}(i), \label{asymptotic:1} \\
&&  \Omega >0 \mbox{ on } \bar{\mathcal{S}}\setminus \{ i\}, \label{asymptotic:2} \\
&& h_{\alpha\beta} = \Omega^2 \Phi_* \tilde{h}_{\alpha\beta}. \label{asymptotic:3}
\end{eqnarray}
\end{subequations}
The last condition shall be, sloppily, written as $h_{\alpha\beta} =
\Omega^2 \tilde{h}_{\alpha\beta}$ ---that is,
$\bar{\mathcal{S}}\setminus \{i\}$ will be identified with
$\tilde{\mathcal{S}}$. Under these assumptions
$(\tilde{\mathcal{S}},\tilde{h}_{\alpha\beta},\tilde{K}_{\alpha\beta})$
will be said to be \emph{asymptotically Euclidean and
regular}. Suitable, punctured neighbourhoods of the point $i$ will be
mapped into the asymptotic end of $\tilde{\mathcal{S}}$.

\medskip 
In the sequel, we will consider initial data sets for the
Einstein field equations which satisfy the boundary conditions
(\ref{asymptotic:1})-(\ref{asymptotic:3}) but have vanishing mass
($m=0$). It follows from the \emph{mass positivity theorem} that if
$m=0$ then either $(\bar{\mathcal{S}},h_{\alpha\beta})$ is data for
the Minkowski spacetime, or the initial data is singular somewhere in
$\tilde{\mathcal{S}}$ ---see e.g. \cite{SchYau79,SchYau81a}. Our
discussion of initial data sets with vanishing mass will be
concentrated on a small neighbourhood $\mathcal{B}_a(i)$ of the point
$i$ where the relevant objects are suitably smooth.

\subsection{Time symmetric initial data sets with vanishing mass}
The construction of purely radiative spacetimes of \cite{Fri88} can be
summarised as follows. Let $V\equiv (\xi^\mu \xi_\mu)^{1/2}$ be the
norm of the timelike Killing vector, $\xi^\mu$, of some static,
asymptotically flat, vacuum spacetime. Assume that the spacetime has
non-vanishing mass. In a static spacetime, there are coordinates
$(t,x^\alpha)$ for which the spacetime metric takes the form
\[
\tilde{g}^{static} = V^2 \mbox{d}t^2 + V^{-2} \tilde{\gamma}_{\alpha\beta} \mbox{d}x^\alpha \mbox{d} x^\beta,
\]
where $\tilde{h}_{\alpha\beta}^{static}\equiv V^{-2}
\tilde{\gamma}_{\alpha\beta}$ denotes the metric of the hypersurfaces
of constant $t$, whereas $\tilde{\gamma}_{\alpha\beta}$ is the metric
of the quotient manifold ---in the case of static spacetimes the
quotient manifold and any surface of constant $t$ can be
identified. Denote by $\tilde{\mathcal{S}}$ one of these (time
symmetric) hypersurfaces of constant $t$. The metric
$\tilde{h}_{\alpha\beta}^{static}$ satisfies on $\tilde{\mathcal{S}}$
the time symmetric Hamiltonian constraint $r[\tilde{h}^{static}]=0$,
whereas $\tilde{\gamma}_{\alpha\beta}$ and $V$ satisfy the 
\emph{vacuum static Einstein field equations}.  Static spacetimes are
usually analysed in terms of the quotient metric
$\tilde{\gamma}_{\alpha\beta}$. The asymptotic behaviour of the
quotient metric $\tilde{\gamma}_{\alpha\beta}$ is best understood in
terms the behaviour of a conformally rescaled quotient metric
$\gamma_{\alpha\beta}\equiv (\Omega^{static})^2
\tilde{\gamma}_{\alpha\beta}$ in a neighbourhood,
$\mathcal{B}_a(i)\subset \bar{\mathcal{S}}$, of the point at infinity
$i$. As before we have
$\bar{\mathcal{S}}=\tilde{\mathcal{S}}\cup\{i\}$. Using the so-called
\emph{conformal static equations} Beig \& Simon \cite{BeiSim81a} have
shown that there is a choice of $\Omega^{static}$ and coordinates for
which $\Omega^{static}$ and $\gamma_{\alpha\beta}$ are analytic in
$\mathcal{B}_a(i)$ and satisfy the boundary conditions
(\ref{asymptotic:1})-(\ref{asymptotic:3}). The crucial observation in
\cite{Fri88} is that the conformally rescaled metric
\[
\tilde{h}_{\alpha\beta} = \frac{1}{2}(1+V)^4 V^{-2}\tilde{h}_{\alpha\beta}^{static},
\]
also satisfies the time symmetric Hamiltonian constraint. The metric
$\tilde{h}_{\alpha\beta}$ is asymptotically Euclidean with vanishing
mass. Let 
\[
\bar{\Omega}\equiv \frac{1}{\sqrt{2}}(1+V)^2 V^{-1}\Omega^{static}, \quad 
\bar{h}_{\alpha\beta}\equiv (\Omega^{static})^2\tilde{h}_{\alpha\beta}=\bar{\Omega}^2 \tilde{h}_{\alpha\beta}^{static}.
\] 
The pair $(\bar{h}_{\alpha\beta},\bar{\Omega})$ satisfies the
time symmetric conformal Hamiltonian constraint, equation
(\ref{conformal_Hamiltonian}), with $\Sigma=0$. Both $\bar{\Omega}$
and $\bar{h}_{\alpha\beta}$ are analytic in $\mathcal{B}_a(i)$ and,
furthermore, they satisfy the boundary conditions
(\ref{asymptotic:1})-(\ref{asymptotic:3}). The pair
$(\bar{h}_{\alpha\beta},\bar{\Omega})$ can be used to construct
initial data for the conformal Einstein field equations of section
\ref{conformal:evolution:equations} ---this data consists essentially
of the value of the rescaled Weyl tensor and the Schouten tensor on
the initial hypersurface. This initial data is obtained out of
algebraic manipulations of the conformal constraint equations.
Although $h_{\alpha\beta}$ and $\Omega$ are given in a gauge which
makes them analytic around $i$, the data for the conformal field
equations need not be. It is a non-trivial result that the data
constructed out of this procedure is indeed analytic around $i$. The
time evolution of this data yields ---using, say, the
Cauchy-Kowalevska theorem--- in a 4-dimensional neighbourhood of $i$
an analytic spacetime metric $g_{\mu\nu}$ and an analytic conformal
factor $\Theta$. 

In the sequel this local existence problem will be solved by a
different method. Under these circumstances, the physical metric,
$\tilde{g}_{\mu\nu}=\Theta^{-2}g_{\mu\nu}$, obtained on the timelike
future $I^+(i)$ of $i$ in this evolution is a solution of the Einstein
vacuum field equations with an analytic structure at past null
infinity, for which the point $i$ represents a regular past timelike
infinity $i^-$. These solutions of the Einstein field equations are
\emph{radiative spacetimes} in the sense discussed in the
introduction.  Note that due to the time reflexion symmetry, the same
holds for the pair $(I^-(i),\tilde{g}_{\mu\nu}) $ where $i$ now
represents the regular future timelike infinity $i^+$. On the
complement of $J^+(i) \cup J^-(i)$ ---these are the points in the
development which are spacelike related to $i$--- one also has a
solution of Einstein's field equations in a neighbourhood of $i$, for
which the point $i$ now represents the point spatial infinity
$i^0$. Definitions of the causal sets $J^\pm$ and $I^\pm$ can be found
in \cite{HawEll73,Wal84}.

\bigskip
The results of \cite{Fri88} relevant for our analysis are presented in the following theorem.

\begin{theorem}
\label{thm:reference:radiative:data}
  Given an analytic solution to the conformal static equations on
$\mathcal{B}_a(i)\subset \bar{\mathcal{S}}$, there exists a solution,
$(\bar{h}_{\alpha\beta},\bar{\Omega})$, to the time symmetric
conformal Hamiltonian constraint on $B_a(i)$ satisfying
\begin{itemize}
\item[(i)] it is asymptotically Euclidean and regular in a neighbourhood $\mathcal{B}_a(i)$,
\item[(ii)] it has vanishing ADM mass,
\item[(iii)] the spinorial fields $\phi_{ABCD}$ and
  $\Theta_{ABCD}$ constructed out of $(\bar{h}_{\alpha\beta},\bar{\Omega})$ are analytic in
  $\mathcal{B}_a(i)$.
\end{itemize}
\end{theorem}

It was long conjectured that given a series of multipole moments
satisfying an appropriate convergence condition, there exists a static
solution of the Einstein equations with precisely those
moments. This conjecture was proved in full generality in \cite{Fri07}
---see also \cite{BaeHer06,Her09} for partial results and an
alternative approach. These results show that the metric, once written
in a certain gauge, is uniquely determined by the multipole
moments. Thus there is an infinite family of static metrics analytic
in a neighbourhood of $i$. This family is parametrised by the
multipole moments. As a consequence of Theorem
\ref{thm:reference:radiative:data} it follows that we have an infinite
family radiative spacetimes constructed in the above mentioned way
that will serve as reference spacetimes for our stability analysis.

\subsection{Construction of radiative reference spacetimes using conformal Gaussian gauges}
In this section we discuss how conformal Gaussian gauge systems can be
used to construct radiative spacetimes out of the Cauchy initial data
provided by Theorem \ref{thm:reference:radiative:data}. The original
construction discussed in \cite{Fri88} makes use of a propagation
system derived from the ``original'' conformal field equations of
\cite{Fri81a,Fri81b,Fri83}. As mentioned before, due to the
analyticity of the setting, existence of a unique analytic solution in
a spacetime neighbourhood of the point $i$ follows from the
Cauchy-Kowalewska theorem ---see e.g. \cite{Eva98}.

\bigskip
Here we follow a different approach. Our intention is to use the
development of the data given by Theorem
\ref{thm:reference:radiative:data} as a reference solution from which
a stability result will be derived. As seen in \cite{LueVal09}, the
discussion of the properties of the conformal boundary of radiative
spacetimes simplifies with the use of a conformal Gaussian system. Thus,
it is natural to discuss the structure and existence of the reference
solution using the same type of gauges. In our approach, existence
of a solution of the propagation equations implied by the extended
conformal field equations and the conformal Gaussian gauge follows
from a variation of a theorem of Kato ---see \cite{Fri86b,Kat75,LueVal09}.

\subsubsection{Setting up conformal Gaussian coordinates}

Let $(\bar{h}_{\alpha\beta},\bar{\Omega})$ on $\bar{\mathcal{S}}$ be
one of the solutions to the time symmetric conformal Hamiltonian
constraint, equation (\ref{conformal_Hamiltonian}), given by Theorem
\ref{thm:reference:radiative:data}. It is assumed that
$\bar{\Sigma}=0$. In what follows all quantities derived from this
initial set and from its time development will be distinguished with
an overbar.  Although it is customary to make use of a conformal
factor $\bar{\Omega}$ which is non-negative, it will be convenient to
consider a non-positive conformal factor on $\bar{\mathcal{S}} $
---the reason for this will be evident in section
\ref{section:hyperboloidal:reference:data}. Accordingly, if one takes
the conformal factor of Theorem \ref{thm:reference:radiative:data}
with an extra factor of $-1$, so that it is negative at least in a
neighbourhood away from $i$. One has that $\bar{\Omega}$ satisfies the
appropriate modification of the boundary conditions
(\ref{asymptotic:1})-(\ref{asymptotic:2}). Namely,
\begin{equation}
\label{asymptotic_behaviour}
\bar{\Omega} < 0 \,\,\mathrm{on}\, \bar{\mathcal{S}}\setminus i , \quad
 \bar{\Omega}(i) = 0  , \quad 
D_{{\cal{A}}}\bar{\Omega}(i) = 0  , \quad 
D_{{\cal{A}}}D_{{\cal{B}}}\bar{\Omega}(i) = 2 h_{{\cal{A}}{\cal{B}}}(i). 
\end{equation}

On a suitably small neighbourhood,
$\mathcal{B}_a(i)\subset\bar{\mathcal{S}}$, we use the coordinates
$x^{\cal{A}}, \, {\cal{A}}=1,2,3$ centered at $i$ and consider the
following initial data for a congruence of conformal geodesics:
\begin{equation}
\label{cg_initial_data}
\bar{\tau}_* = 0 , \quad \dot{x}^\mu = \bar{n}^\mu, \quad \bar{\Theta}_* = \bar{\Omega} , \quad \dot{\bar{\Theta}}_* =0, \quad \bar{d}_* \equiv \bar{\Theta}_* \bar{b}_* =(\mathrm{d}\bar{\Theta})_*.
\end{equation}
The coordinates $x^{\cal{A}}$ are extended off $\bar{\mathcal{S}}$ by
dragging along the congruence of conformal geodesics to obtain
conformal Gaussian coordinates as discussed in section
\ref{section:conformal:gaussian:gauges}. It follows that $\bar{d}_*=0
$ at $i$ and using $\bar{d}_a(\bar{\tau})=\bar{d}_a(0)$, we get
$\bar{d}_a = D_a(\bar{\Omega})$. The constraint
(\ref{Theta:constraint}) applied to our situation gives the initial
data for $\ddot{\bar{\Theta}} $ on $\mathcal{B}_a\setminus i $,
which must hence be positive there. The initial value at $i$ is
obtained taking the corresponding limit. Making use of normal
coordinates centred in $i$, one finds that since $\bar{\Omega} =
\mathcal{O}(\vert x \vert^2) $, we have $h^\sharp(D\bar{\Omega},
D\bar{\Omega}) = \mathcal{O}(\vert x \vert^2)$ and thus
$\ddot{\bar{\Theta}} = \mathcal{O}(1)$. Hence $\ddot{\bar{\Theta}}_* >
0$ everywhere on $\bar{\mathcal{S}} $. It follows from equation
(\ref{general:Theta}) that along each conformal geodesic the conformal
factor $\bar{\Theta}$ is given by
\begin{equation}
\label{special:Theta}
\bar{\Theta}(\bar{\tau}) = \bar{\Omega} + \frac{1}{2} \ddot{\bar{\Theta}}_* \bar{\tau}^2 = \bar{\Omega}\left(1 - \frac{\bar{\tau}^2}{\bar{\omega}^2} \right)
\end{equation}
where 
\[
\bar{\omega} \equiv  \sqrt{\frac{2\bar{\Omega}}{\ddot{\bar{\Theta}}_*}}, \quad \bar{\omega} (i)=0.
\]
Define now the conformal boundary, $\bar{\scri}$, in a natural way as
the locus of points in the development of the data on
$\bar{\mathcal{S}}$ for which $\bar{\Theta}=0$ and
$\mbox{d}\bar{\Theta}\neq 0.$ It is easy to see that a conformal
geodesic with data given by (\ref{cg_initial_data}) passes through $\bar{\mathscr{I}}$
whenever $\bar{\tau} =\pm \bar{\omega}$. Note that every conformal geodesic, except for the
one passing through $i$, intersects $\bar{\mathscr{I}}$ twice, once to
the past of $\bar{\mathcal{S}}$ and once to the future.

\bigskip
\textbf{Remark.} We shall always work on a neighbourhood
$\mathcal{B}_a(i)\subset \bar{\mathcal{S}}$ which is assumed to be
sufficiently small to ensure $h^{\alpha\beta}D_\alpha \bar{\Omega} D_\beta \bar{\Omega}\ne 0 $ in a
neighbourhood of $i$.


\subsubsection{Existence and conformal structure of the reference spacetime} \label{refST}

We discuss now the existence of solutions to the propagation system
(\ref{sh:system}) with data prescribed on $\mathcal{B}_a(i)\subset
\bar{\mathcal{S}}$. The initial data for this initial value problem is
obtained through the conformal constraint equations on
$\bar{\mathcal{S}}$ \cite{Fri83,Fri88} and satisfies on
$\mathcal{B}_a(i)\subset \bar{\mathcal{S}}$
\begin{subequations}
\begin{eqnarray}
&& e^0_{AB}=0, \label{reference:data1} \\
&& e^\er_{AB}= \sigma^\er_{AB}, \\
&& f_{AB}=0, \\
&& \xi_{ABCD}= \xi_{ABCD}[\bar{h}],  \\
&& \chi_{(AB)CD}=0, \\
&& \Theta_{ABCD} = -\frac{1}{\bar{\Omega}} D_{(AB} D_{CD)} \bar{\Omega} + \frac{1}{12}h_{ABCD}, \\
&& \phi_{ABCD} = \frac{1}{\bar{\Omega}^2} D_{(AB} D_{CD)} \bar{\Omega} + \frac{1}{\bar{\Omega}} s_{ABCD}, \\
&& \eta = 0, \\
&& \eta_{AB} = \sigma^\er_{AB} ,
\label{reference:data9}
\end{eqnarray}
\end{subequations} 
where $h_{ABCD}\equiv -\epsilon_{A(C} \epsilon_{D)B}$ and
$\sigma^\er_{AB}$ are the spatial Infeld-van der Waerden symbols. The
spinorial fields $\xi_{ABCD}[\bar{h}]$ and $s_{ABCD}$ are the
connection coefficients and the trace-free Ricci curvature of the
3-metric $\bar{h}_{\alpha\beta}$.

\bigskip
Since $\mathcal{B}_a(i)$ is assumed to be a simply connected
neighbourhood of $i$, we can find a diffeomorphism $\phi_{\mathcal{B}}
$ onto $\mathcal{U} \subset \Sphere^3 $. We extend the initial data to
$\Sphere^3 $ in the following way. The initial values for
\[
e^\es_{AB}, \quad  f_{AB}, \quad \chi_{(AB)CD}, \quad  \eta, \quad \eta_{AB}
\]
are constant and are extended with their respective values to all of
$\Sphere^3$. For
\begin{equation}
\label{preextensionlabel}
\bar{\Theta}_*, \quad \bar{d}_*, \quad \xi_{ABCD}, \quad \Theta_{ABCD}, \quad  \phi_{ABCD}
\end{equation}
we use the linear extension operator $E$ given in section \ref{section extending initial data} to obtain fields
defined over $\Sphere^3$. The extended fields will be denoted on $\Sphere^3$ in the same way as in (\ref{preextensionlabel}).

\bigskip 
The vectorial unknown 
\[
u(\tau, x) \equiv \big(\mbox{Re}(\upsilon)(\tau, x),\mbox{Im}(\upsilon)(\tau, x),\mbox{Re}(\phi)(\tau, x),\mbox{Im}(\phi)(\tau, x) \big)
\]
is viewed as a function of $\tau $ with values in the Sobolev space
$H^s(\Sphere^3, \Real^N)$. Let $u_0(x)=u(0,x)$ denote the extended initial
data on $\Sphere^3$ as described in the previous paragraphs. For
$\delta\in\Real$, $m\in\Natural$, $m\geq 2$, define set
\begin{equation}
\label{positivedefiniteness:condition}
D_\delta^m=\left\{w\in H^m(\Sphere^3,\Real^N) \; |\; (z,A^0(w)z)>\delta(z,z), \forall z\in\Real^N    \right\},
\end{equation}
where $A^0(w)$ denotes the matrix valued function defined by the
symmetric hyperbolic system (\ref{sh:system}), and $(\cdot,\cdot)$ is
the standard scalar product on $\Real^N$. It can be verified that
\[
A^0(u_0) =\mbox{diag}\left(1,\dots,1,\frac{1}{\sqrt{2}},\sqrt{2},\sqrt{2},\sqrt{2},\frac{1}{\sqrt{2}},\frac{1}{\sqrt{2}},\sqrt{2},\sqrt{2},\sqrt{2},\frac{1}{\sqrt{2}}\right).
\]
and hence $u_0 \in D_\delta^m$ for $0<\delta <1/\sqrt{2}$. 

\bigskip
One has the following local existence result.

\begin{theorem}
\label{Katopart1}
  Suppose $m\geq 4$. There exists a $\bar{T}_0>0$, a subset
$D_{\bar{T}_0}\subset D_\delta^m$ and a unique solution $u(\tau)$ of
equation (\ref{sh:system}) defined on $[-\bar{T}_0, \bar{T}_0]$ with
initial data $u_0$ on $\bar{\mathcal{S}} \simeq \Sphere^3$ and such that
  \[
  u\in C(-\bar{T}_0,\bar{T}_0;D_{\bar{T}_0})\cap C^1\big(0,T; H^{m-1}(\Sphere^3,\Real^N)\big).
  \]
  Furthermore, $u \in C^\infty([-\bar{T}_0, \bar{T}_0] \times \Sphere^3)$ and $\bar{T}_0$ can be chosen such that $u(\tau)$ has non-vanishing Jacobi fields $\eta_{AB} $, so that the solution is free of conjugate points in $[-\bar{T}_0,\bar{T}_0]$.
\end{theorem}

\noindent
\textbf{Proof.} The first part of the theorem follows from the
generalisation given in \cite{Fri86b} ---see also \cite{LueVal09}---
of Kato's existence and stability result for quasilinear symmetric
hyperbolic systems \cite{Kat75}. For the Jacobi fields we observe that
$\eta_{AB} $ satisfies an ordinary differential equation along the
curves of the congruence. Hence there is a minimum interval for which
the fields do not vanish. The lack of conjugate points then follows
from our earlier discussion. \hfill $\square$

\bigskip
\noindent
\textbf{Remark 1.} It follows that on $(-\bar{T}_0,\bar{T}_0)$ the
solution, as given by the theorem, is of class 
\[
H^m\big((-\bar{T}_0,\bar{T}_0)\times \Sphere^3\big)
\subset C^{m-2}\big((-\bar{T}_0,\bar{T}_0)\times \Sphere^3\big).
\]

\bigskip
\noindent
\textbf{Remark 2.} Let $\mathring{\mathcal{M}} \equiv [-\bar{T}_0 ,
\bar{T}_0 ] \times \Sphere^3$ and denote the metric on
$\mathring{\mathcal{M}} $ by $\mathring{g}_{\mu\nu} $. The spacetime
$(\mathring{\mathcal{M}}, \mathring{g}_{\mu\nu} ) $ will be used as
the reference spacetime for our stability analysis later on. Let
$D(\mathcal{B}_a(i))$ denote the domain of dependence of
$\mathcal{B}_a(i)$ ---see e.g. \cite{HawEll73,Wal84} for
definitions. The value of $\bar{T}_0$ can be chosen small enough so that
\begin{equation}
\mathcal{M}  \equiv  \mathring{\mathcal{M}}  \cap J^-(i) \subset D(\mathcal{B}_a(i)).
\end{equation}

\begin{figure}[t]
\centerline{\includegraphics[width=.6\textwidth]{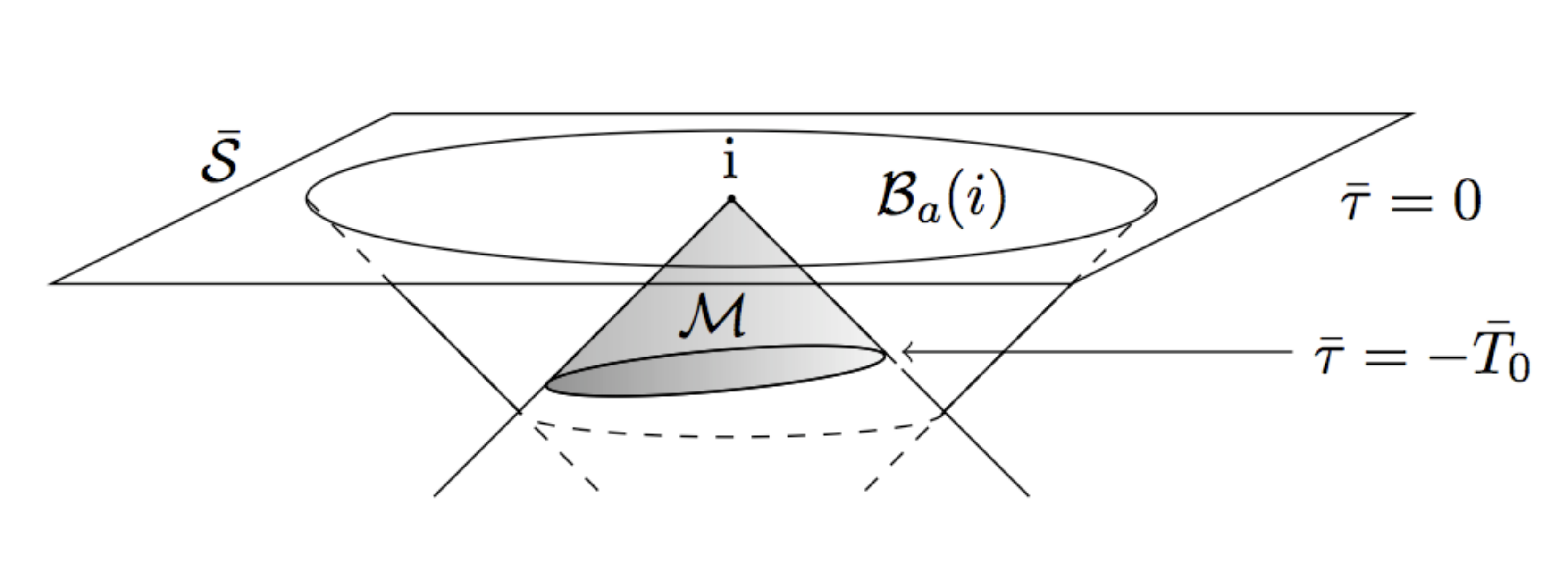}}
\caption{A conformal representation of the reference solution obtained as the development of
Euclidean initial data with vanishing mass on $\bar{\mathcal{S}}$. The shaded region represents the physical spacetime.}
\end{figure}

\bigskip
The finite reduction of $\bar{T}_0$ is not a
problem in our scheme as any $\bar{T}_0>0$ corresponds to an infinite
time interval in the physical description. As it will be seen in the
sequel, it may be necessary to further reduce $\bar{T}_0$ by a finite
(non-negative) amount.

\bigskip
The spacetime $(\mathcal{M}\setminus \bar{\mathscr{I}},
\mathring{g}_{\mu\nu} ) $ is conformally related to a vacuum
spacetime, $(\mathcal{M}\setminus \bar{\mathscr{I}},
\Theta^{-2}\mathring{g}_{\mu\nu})$, with vanishing cosmological
constant. The spacetime $(\mathcal{M}\setminus \bar{\mathscr{I}},
\Theta^{-2} \mathring{g}_{\mu\nu} ) $ is a radiative spacetime as the
conformal metric $g_{\mu\nu}$ and associated fields extend smoothly
through the conformal boundary $\bar{\mathscr{I}}$ and furthermore,
there exists a point $i^+=(0,i)\in \{0\} \times \bar{\mathcal{S}}
\subset \mathring{\mathcal{M}}$ satisfying (by construction) the
conditions to be a \emph{future timelike infinity} ---cfr. the
discussion in the introduction. As a result of Lemma
\ref{lemma:propagation constraints} it follows that the conformal
constraints are only propagated in
$D(\mathcal{B}_a(i))$. Consequently, the region $ [-\bar{T}_0 ,
\bar{T}_0 ] \times \Sphere^3\setminus D(\mathcal{B}_a(i))$ is of no
physical interest as it does not describe a vacuum spacetime. It is
nevertheless relevant to use the whole of
$(\mathring{\mathcal{M}},\mathring{g}_{\mu\nu} )$ for the correct
formulation of the stability part of our work. In the following, in a
slight abuse of language, we will refer to $(\mathcal{M},
\mathring{g}_{\mu\nu} )$ as \emph{the purely radiative reference
solution}.

\subsubsection{Hyperboloidal data for the radiative reference spacetime}
\label{section:hyperboloidal:reference:data}

The conformal affine parameter $\bar{\tau}$ defines, in a natural way, 
a foliation of the manifold $\mathring{\mathcal{M}}$. Let
$\mathcal{S}_{\bar{\tau}}$ denote the surfaces of constant
$\bar{\tau}$. For fixed $\bar{\tau}$ one has that $\mathcal{S}_{\bar{\tau}}$ is
diffeomorphic to $\Sphere^3$.  Let $\bar{\tau}_0 \in (0,\bar{T}_0)$
and define
\[
\mathcal{S}_0 \equiv \{ - \bar{\tau}_0\}\times \Sphere^3 , \quad \mathcal{Z} \equiv \{p \in
\mathcal{S}_0 \vert \bar{\Theta} = 0\}.
\]
The set $\mathcal{S}_0$ intersects null infinity in a
\emph{hyperboloidal way}. More precisely, $\mathcal{Z}$ divides
$\mathcal{S}_0$ into two regions: one where $\bar{\Theta}<0$ and one
where $\bar{\Theta}>0$. The latter contains the origin and will be
denoted by $\mathring{\mathcal{H}}$. Hence
\[
\mathring{\mathcal{H}} \equiv \{ p\in \mathcal{S}_0 \vert \bar{\Theta} >0\}.
\]
The latter definition justifies the choice of a negative conformal
 factor $\bar{\Omega}$ made in the previous sections.  We rewrite the
 conformal parameter $\bar{\tau}$ and the conformal factor of the
 reference solution given by equation (\ref{special:Theta}), so that
 their initial surface is $\mathcal{S}_0$. Define
\[
\tau\equiv \bar{\tau} + \bar{\tau}_0, \quad \mathring{\Theta}(\tau) \equiv
\bar{\Theta}(\tau - \bar{\tau}_0)
\] 
so that $\tau =0 $ on $\mathcal{S}_0$ and
\begin{eqnarray*}
&&\mathring{\Theta} (\tau) = \bar{\Omega} \left( 1-\frac{\bar{\tau}^2_0}{\bar{\omega}^2} \right) - \ddot{\bar{\Theta}}_* \bar{\tau}_0 \tau + \frac{1}{2} \tau^2 \ddot{\bar{\Theta}}_* \\
&& \phantom{\Theta (\tau)}= \bar{\Omega} \left( \left( 1-\frac{\bar{\tau}^2_0}{\bar{\omega}^2} \right) + 2\frac{\bar{\tau}_0}{\bar{\omega}^2}\tau - \frac{1}{\bar{\omega}^2}\tau^2 \right).
\end{eqnarray*}
The initial value of $\mathring{\Theta}$ on $\mathcal{S}_0$ will be denoted
 by $\mathring{\Omega}$. Recalling that $\bar{\tau}_\pm(x^\mathcal{A}) =
\pm\bar{\omega}(x^\mathcal{A})$ gives the location of null infinity, we can see
that
\begin{eqnarray*}
\mathring{\Omega} > 0 &\iff & \vert\bar{\omega}\vert > \vert\bar{\tau}_0\vert ,\nonumber \\
\mathring{\Omega} = 0 &\iff & \vert\bar{\omega}\vert = \vert\bar{\tau}_0\vert , \\
\mathring{\Omega} < 0 &\iff & \vert\bar{\omega}\vert < \vert\bar{\tau}_0\vert , \nonumber 
\end{eqnarray*}
and hence $\mathring{\Omega}$ is the correct boundary defining
function on $\mathcal{S}_0$ with $\mathcal{Z} = \{ p \in \mathcal{S}_0
\vert \mathring{\Omega}=0 \}= \partial \mathring{\mathcal{H}}$. One
can briefly verify that the new conformal factor $\mathring{\Theta}$
gives the correct location of null infinity. If $\bar{\Omega} \ne 0 $
we get that $\mathring{\Theta} = 0$ whenever
\[
\tau =  \frac{ \displaystyle  - \frac{2\bar{\tau}_0 }{\bar{\omega}^2} \pm \sqrt{\frac{4\bar{\tau}_0^2}{\bar{\omega}^4} + \frac{4}{\bar{\omega}^2} \left( 1 - \frac{\bar{\tau}^2_0}{\bar{\omega}^2}     \right)}   }{\displaystyle -\left( \frac{2}{\bar{\omega}^2}\right)}
= \bar{\tau}_0 \pm \bar{\omega}.
\]
Along the unique conformal geodesic $\gamma$ passing through the point
$N$ on $\Sphere^3$ at all times, we have $\bar{\Omega} = 0 $ and
\[
\mathring{\Theta} = \frac{1}{2}\ddot{\bar{\Theta}}_*(\tau - \bar{\tau}_0)^2.
\]
Thus, along $\gamma$ one has that $\mathring{\Theta} = 0 $ only when
$\tau= \bar{\tau}_0$. This corresponds to the point $i$ in the Cauchy
hypersurface $\bar{S}$.

\bigskip
In what follows, let
\[
\mathring{u}(\tau,x)\equiv u(\tau-\bar{\tau}_0,x).
\]
For consistency, we shall denote entries of $\mathring{u}$, as defined
above, also with $\mathring{\phantom{Q}}$. The entries of
$\mathring{u}$ imply on $\mathcal{S}_0$ data for the symmetric
hyperbolic system (\ref{sh:system}). We shall write
$\mathring{u}_0(x)\equiv \mathring{u}(0,x)$. The data $\mathring{u}_0$
are not truly hyperboloidal data. The reason for this is the
following. The frame vector $\mathring{e}_0$ which on
$\bar{\mathcal{S}}$ is normal to this surface, ceases to be normal to
the surfaces $\mathcal{S}_{\bar{\tau}}$, of constant $\bar{\tau}$ if
$\bar{\tau}\neq 0$. This is a feature that differentiates conformal
geodesics from standard geodesics. This phenomenon manifests itself
in the fact that the components $\mathring{e}^0_{AB}$ of the space
spinor decomposition of $\mathring{e}_0$ according to formulae
(\ref{space:spinor:decomposition:1}) and
(\ref{space:spinor:decomposition:2}), which are
zero on $\bar{\mathcal{S}}$, become, in general non-zero off $\bar{\mathcal{S}}$
\footnote{Note, however, that in the case of the conformal Minkowski
spacetime in the form discussed in \cite{LueVal09}, tangent vectors to
the congruence of conformal geodesics are always normal to the surfaces
of constant affine conformal parameter.}. As a consequence, the
spatial vectors of the frame $\mathring{e}_a$, $a=1,2,3$ which span
the tangent bundle to $\bar{\mathcal{S}}$, $T\bar{\mathcal{S}}$, in general, pick
up components off the tangent bundle
$T\mathcal{S}_{\bar{\tau}}$. Accordingly, the data $\mathring{u}_0$,
which is expressed in terms of the spatial frame $\mathring{e}_a$
($\mathring{e}_{AB}$) is not intrinsic to $\mathcal{S}_0$. We shall
call this type of data \emph{tilted hyperboloidal data}.

\bigskip
In order to relate the tilted data $\mathring{u}_0$ to proper
hyperboloidal data one proceeds as follows.  Denote the normal vector
field to $\mathcal{S}_0$, the 3-metric induced by
$\mathring{g}_{\mu\nu}$ and the second fundamental form of
$\mathcal{S}_0$ by $n^\mu$, $\mathring{h}_{\alpha\beta}$ and
$\mathring{K}_{\alpha\beta}$ respectively. Note that $\mathring{K}_{\alpha\beta}$ and
$\chi_{\alpha\beta}$ are two conceptually different quantities --- the
former is related to the hypersurface, the later to the conformal
geodesic congruence. Choose a $\mathring{h}$-orthonormal frame
$\underline{\mathring{e}}_a, \; a=1,2,3 $ and set
$\underline{\mathring{e}}_0 = n$, so that $\underline{\mathring{e}}_k,
\; k=0,1,2,3 $, forms a $\mathring{g}$-orthonormal frame on
$\mathcal{S}_0$. The frames $\{v, \mathring{e}_a \}$ and $\{n,
\underline{\mathring{e}}_a \} $ are related by a Lorentz
transformation $\updn{\Lambda}{i}{j}=\updn{\Lambda}{i}{j}(x)$, which
will be regarded as a matrix valued function over $\Sphere^3$, as
follows
\begin{subequations}
\begin{eqnarray}
&& \updn{\Lambda}{i}{j}: \mathring{e}_i \rightarrow \underline{\mathring{e}}_j, \quad \updn{\Lambda}{i}{j} \dnup{\Lambda}{k}{j}=\updn{\delta}{i}{k}, \label{Lorentz:transformation:1} \\
&& \underline{\mathring{e}}_j = \updn{\Lambda}{k}{j}\mathring{e}_k, \label{Lorentz:transformation:2} \\
&& \mathring{e}_j =\dnup{\Lambda}{j}{k}\underline{\mathring{e}}_k. \label{Lorentz:transformation:3}
\end{eqnarray}
\end{subequations}
The fields 
${\mathring{f}}_i$,
$\dnupdn{{\mathring{\Gamma}}}{i}{j}{k}$,
${\mathring{L}}_{ij}$, ${\mathring{d}}_{ijkl}$
are the frame components of spacetime quantities and the Lorentz transformation induces transformed components
$\underline{\mathring{f}}_i$,
$\dnupdn{\underline{\mathring{\Gamma}}}{i}{j}{k}$,
$\underline{\mathring{L}}_{ij}$, $\underline{\mathring{d}}_{ijkl}$ in
the canonical way. For example
\[
\underline{\mathring{d}}_{ijkl} = \updn{\Lambda}{m}{i} \updn{\Lambda}{n}{j} \updn{\Lambda}{p}{k} \updn{\Lambda}{q}{l} \mathring{d}_{mnpq}.
\]
The SL(2,\Complex)-spinorial counterparts of the above tensors can be
obtained by suitable contractions with the constant Infeld-van der
Waerden symbols. These, in turn, can be space-spinor split with respect
to $\underline{\mathring{\tau}}^\mu\equiv \sqrt{2} \underline{\mathring{e}}_0$ to obtain \emph{proper hyperboloidal data} on
$\mathring{\mathcal{H}}$.

\section{Radiative spacetimes from perturbed hyperboloidal data}
\label{section:perturbed:radiative:spacetime}

Given one of the radiative spacetimes discussed in section
\ref{section:reference:radiative:spacetime}, it is natural to ask
whether it is possible to establish for these spacetimes an analogue
of the semiglobal stability results for the Minkowski spacetime
discussed in \cite{Fri88,LueVal09}. In this section we establish such
a result. This result generalises the aforementioned stability results
to the case where the reference solutions are not flat.

\subsection{On hyperboloidal data}
\label{section:hyperboloidal:data}

In order to discuss appropriately the notion of closeness of two
hyperboloidal initial data sets, we make use of the notion of 
\emph{limit of spacetimes} as discussed in e.g \cite{Ger69}. Assume
that one has a smooth family of hyperboloidal initial data sets
$(\mathcal{H}^\varepsilon,h_{\alpha\beta}^\varepsilon,K_{\alpha\beta}^\varepsilon,\Omega^\varepsilon,\Sigma^\varepsilon)$,
parametrised by $\varepsilon\geq 0$, such that
\[
(\mathcal{H}^0,h_{\alpha\beta}^0,K_{\alpha\beta}^0,\Omega^0,\Sigma^0)= (\mathring{\mathcal{H}},\mathring{h}_{\alpha\beta},\mathring{K}_{\alpha\beta},\mathring{\Omega},\mathring{\Sigma}), 
\]
where $\mathcal{H}^\varepsilon$ are simply connected 3-dimensional
manifolds with boundary $\partial \mathcal{H}^\varepsilon$. The fields
$(h_{\alpha\beta}^\varepsilon,K_{\alpha\beta}^\varepsilon,\Omega^\varepsilon,\Sigma^\varepsilon)$
satisfy the vacuum constraints, equations
(\ref{conformal_Hamiltonian}) and (\ref{conformal_Momentum}), with
\[
\Omega^\varepsilon=0, \quad  \mbox{d}\Omega^\varepsilon \neq
0 \mbox{  on }\partial \mathcal{H}^\varepsilon, \quad  \Omega^\varepsilon>0 \mbox{ on the interior of } \mathcal{H}^\varepsilon.
\]
As the manifolds $\mathcal{H}^\varepsilon$ are taken to be simply
connected, there exists a family of diffeomorphisms
\[
\phi^\varepsilon: \mathcal{H}^\varepsilon \rightarrow \mathcal{U}\subset \Sphere^3, \quad \phi^\varepsilon(\partial\mathcal{H}^\varepsilon)= \partial \mathcal{U},
\]
with $\mathcal{U}$ as in section \ref{coordinates:S3}. In particular, 
\[
\Phi^\varepsilon\equiv \mathring{\phi}^{-1} \circ \phi^\varepsilon: \mathcal{H}^\varepsilon \rightarrow \mathring{\mathcal{H}} \quad \mbox{ with } \quad \mathring{\phi}\equiv \phi^0,
\]
is a smooth family of diffeomorphisms between
$\mathcal{H}^\varepsilon$ and $\mathring{\mathcal{H}}$, such that
$\Phi^0$ is the identity map. Note that the locus of points for which
$\Omega^\varepsilon=0$ coincides as sets on $\Sphere^3$, by
construction, for all $\varepsilon\geq 0$.

\bigskip
Let $\{ \underline{e}^\varepsilon_a \}$, $a=1,\,2,\,3$ denote a smooth
family of $h^\varepsilon$-orthonormal frames such that
\[
\underline{e}^{0}_a = \underline{\mathring{e}}_a, \quad a=1, \,2,\, 3,
\] 
with $\underline{\mathring{e}}_a$ as given in the end of section
\ref{section:hyperboloidal:reference:data}. The Levi-Civita connection
$D^\varepsilon$ of the metric $h^\varepsilon_{\alpha\beta}$,
together with the frame $ \underline{e}^\varepsilon_a$, imply the
3-dimensional Ricci tensor $r[h^\varepsilon]_{\alpha\beta}$ 
and spin connection coefficients
$\dnupdn{\underline{\gamma}}{a}{b}{c}$, as well as their spinorial
counterparts $\underline{r}_{ABCD}^\varepsilon $ and
$\underline{\xi}_{ABCD}^\varepsilon$. From these quantities and the
tensor $K_{\alpha\beta}^\varepsilon$ one can determine the
$\underline{e}_a$-components of the Schouten tensor and 
$\underline{e}_0$-electric and $\underline{e}_0$-magnetic parts of the
rescaled Weyl tensor for the spacetime on $\mathcal{H}^\varepsilon$
---see e.g. \cite{Fri83,Fri86b}. Using the frame $\underline{e}_a$ and
$\underline{\tau}^\mu \equiv \sqrt{2}\underline{e}_0 $, their
spinorial counterparts can be readily calculated.  This procedure
produces proper hyperboloidal data ---in the sense discussed in
section \ref{section:hyperboloidal:reference:data}. We note that so
far the data is only given on $\mathcal{H}^\varepsilon $, respectively,
$\phi^\varepsilon ( \mathcal{H}^\varepsilon ) = \mathcal{U}\subset
\Sphere^3$.  We use the extension operator $E$ given in section
\ref{section extending initial data} to extend the proper
hyperboloidal data to the whole of $\Sphere^3$, bearing in mind that
outside $\mathcal{U}$ the data may violate the vacuum constraints
(\ref{conformal_Hamiltonian}) and (\ref{conformal_Momentum}).

\bigskip
In the following, the parameter $\varepsilon$ will be dropped from the
expressions, in order to ease the presentation.

\subsubsection{Local existence for hyperboloidal data}

From studying the proper hyperboloidal data on the hypersurface
$\mathcal{S}_0$ in the reference solution one can see that this data
does not include all the quantities that are needed for the tilted
hyperboloidal data used for (\ref{sh:system}). We are missing the data
for $e^{\bar{s}}_{AB}$, $f_{AB}$, $\xi_{ABCD}$, $\chi_{ABCD} $.

In order to consider the construction of tilted hyperboloidal data for
the conformal propagation equations we shall first develop a small
part of the perturbed spacetime. For this we need a local existence
result for the development of data constructed using an auxilary
congruence of conformal geodesics which departs the initial
hyperboloid orthogonally. The information thus obtained complements
the proper hyperboloidal data constructed above.

For the construction of the auxilary congruence we use the following
initial setup.  Let $\dot{x} = n$, $e_k\vert_{\mathcal{S}} =
\underline{e}_k $, $\underline{\xi}_{ABCD}\vert_{\mathcal{S}} =
\xi_{ABCD}[\bar{h}] $, $\underline{\chi}_{(AB)CD}\vert_{\mathcal{S}} =
K_{ABCD} $. As long as the data for $\underline{f}_{AB}$ is smooth and
satisfies $\langle \underline{f},n \rangle =0 $,
i.e. $\underline{f}_{AB}=\underline{f}_{(AB)}$ it can be chosen
arbitrarily, since we are only interested in local existience in some
small neighbourhood of the initial surface.  In particular, we can set
$\underline{f}_{AB}=0$ again.

We can now proceed to our local existence result for the unknown
\[
\underline{u}=(\underline{e}_{AB}^0,\underline{e}^\er_{AB},\underline{f}_{AB}, \underline{\xi}_{ABCD}, \underline{\chi}_{(AB)CD},\underline{\Theta}_{ABCD},\underline{\phi}_{ABCD}, \underline{\eta},\underline{\eta}_{AB}).
\]
The underline in the above spinors indicates that they are expressed
with respect to the frame $\underline{e}_a$.
The proposition below is essentially very similar to Theorem \ref{Katopart1}. 
This is not a stability result, but purely a local existence 
for the perturbed data. For this we define $\underline{D}^m_\delta$ analogously 
to (\ref{positivedefiniteness:condition}).

\begin{proposition}
\label{lemma:local existence}
  Suppose $m\geq 4$. There exists a $\underline{T}_0>0$, a subset
$\underline{D}_{\underline{T}_0}\subset \underline{D}_\delta^m$ 
and a unique solution $\underline{u}(\tau)$ of
equation (\ref{sh:system}) defined on $[-\underline{T}_0, \underline{T}_0]$ with
initial data $\underline{u}_0$ on $\mathcal{S} \simeq \Sphere^3$ and such that
  \[
  u\in C(-\underline{T}_0,\underline{T}_0;D_{\underline{T}_0})\cap C^1\big(0,T; H^{m-1}(\Sphere^3,\Real^N)\big).
  \]
  Furthermore, $\underline{u} \in C^\infty([-\underline{T}_0,
\underline{T}_0] \times \Sphere^3)$ and $\underline{u}(\tau)$ has
non-vanishing Jacobi fields $\underline{\eta}_{AB} $, so that the
solution is free of conjugate points in
$[-\underline{T}_0,\underline{T}_0]$.
\end{proposition}

\textbf{Remark:} The local existence theorem clearly applies to
$(\mathring{\mathcal{H}},\mathring{h}_{\alpha\beta},\mathring{K}_{\alpha\beta},\mathring{\Omega},\mathring{\Sigma})$
and the resulting spacetime coincides locally with the reference
spacetime $(\mathring{\mathcal{M}}, \mathring{g}_{\mu\nu})$. One could
then consider the derivation of a result similar to the stability of
Minkowski in \cite{LueVal09}. However it should be noted that there is
no guarantee that this setup covers $i^+$ or $\mathscr{I}^+ $ of
$(\mathring{\mathcal{M}}, \mathring{g}_{\mu\nu})$ and its
perturbations. Thus a semiglobal existence and stability result can
not be guaranteed.

\subsubsection{Construction of tilted hyperboloidal data}

As a result of Proposition \ref{lemma:local existence} we have a
spacetime slab $\mathcal{N} \equiv [-\underline{T}_0, \underline{T}_0]
\times \Sphere^3$ with metric $ g_{\mu\nu}$ which describes a vacuum
spacetime on $\underline{\mathcal{M}} \equiv \mathcal{N} \cap
D(\mathcal{H})$.  For sufficiently small $\varepsilon$, one can regard
the 3-manifold $\mathcal{S}$ as a hypersurface of a spacetime
$(\mathcal{N},g_{\mu\nu})$. On this hypersurface we now want to give
tilted hyperboloidal data for (\ref{sh:system}).  For this we will
evolve a conformal geodesic congruence, that departs $\mathcal{S}$ in
the same way as in the reference solution. Thus we can locally obtain
all the components of the unknown $u$ on $\mathcal {S}$ and use this
as initial data $u_0$ for (\ref{sh:system}).

\bigskip
We know fix the initial data for the frame and the congruence.
As before, let $n$ denote the unit normal to $\mathcal{S}$. 
Then $\{\underline{e}_i\}\equiv\{n,\underline{e}_a\}$
is an orthonormal tetrad on $\mathcal{S}$. From $ \underline{e}_j $ we
construct a tilted orthonormal frame $ e_i $ on $T \mathcal{N}\vert_{\mathcal{S}}$ via
\[
e_i = \dnup{\Lambda}{i}{j} \underline{e}_j,
\]
where $\dnup{\Lambda}{j}{i}$ is the \emph{same} matrix valued function
associated to the Lorentz transformation via
(\ref{Lorentz:transformation:1})-(\ref{Lorentz:transformation:3}). It
follows then that the timelike vector $e_0$ has the same projections
on the orthonormal frame $\underline{e}_j$ as $\mathring{e}_0$ has on
$\underline{\mathring{e}}_j$.  That is $g(e_0,
\underline{e}_j)=\mathring{g}(\mathring{e}_0,
\underline{\mathring{e}}_j) $. Note that $\mathring{e}_0$ is by
construction tangent to a congruence of conformal geodesics.  For
discussions of the development of the \emph{perturbed initial data}
$(\mathcal{H},\gamma_{\alpha\beta},K_{\alpha\beta},\Omega,\Sigma)$
which go beyond local existence, it will be important to consider a
congruence of conformal geodesics that departs the hypersurface
$\mathcal{H}$ in the same way as the congruence of conformal geodesics in
the reference spacetime arrives at $\mathring{\mathcal{H}}$. For this
reason we will use $ \dot{x} = e_0 $.

\begin{figure}[t]
\centering
 \includegraphics[width=.5\textwidth]{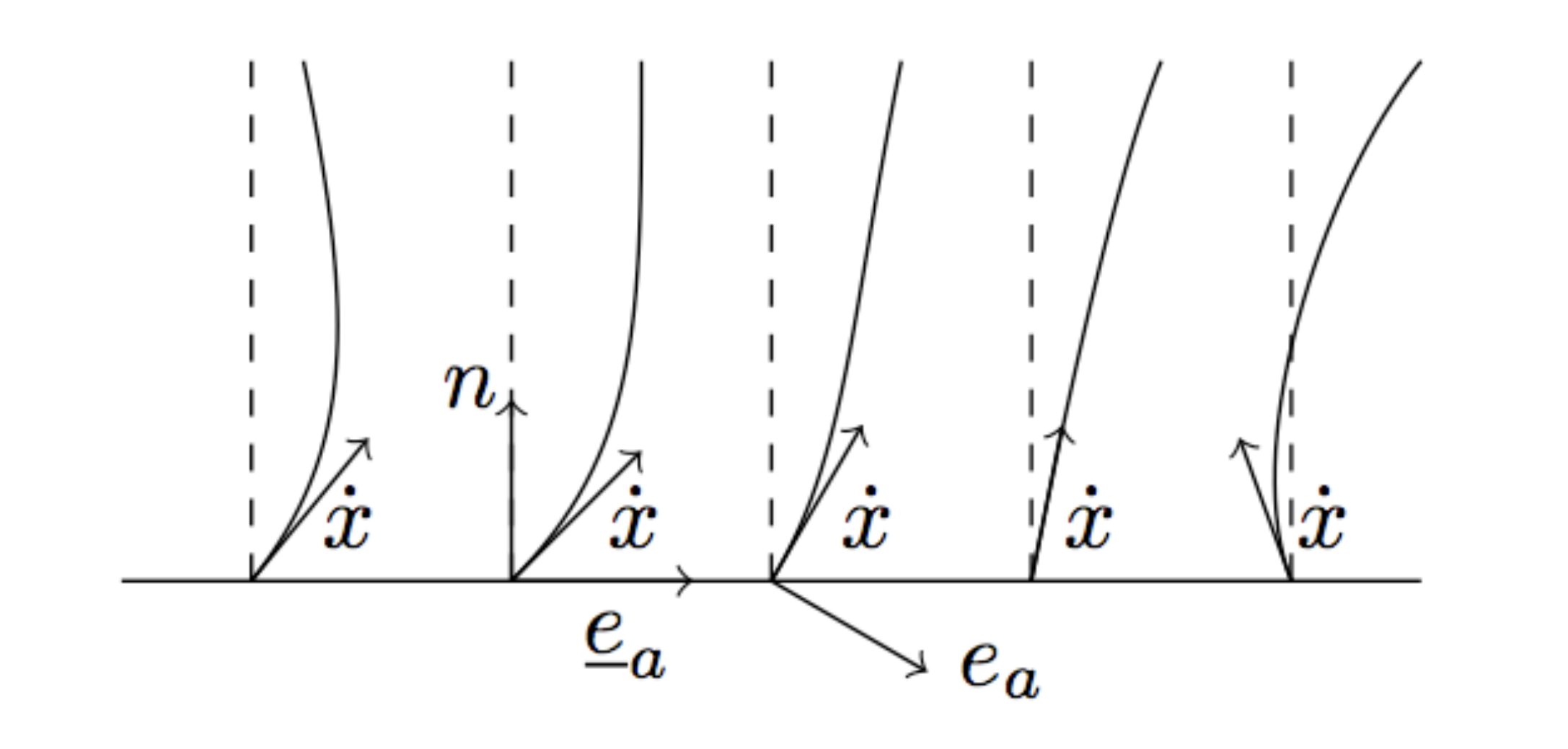}
\caption{The auxilary congruence ($-\,-\,-$) normal to $\mathcal{H}$ and the tilted congruence (---) used to evolve the perturbed initial data.}
\end{figure}

\bigskip
As in the case of the reference spacetime
$(\mathring{\mathcal{M}},\mathring{g}_{\mu\nu})$, we want to cover the
development of the initial data set
$(\mathcal{H},h_{\alpha\beta},K_{\alpha\beta},\Omega,\Sigma)$ with a
conformal geodesic coordinate system. The formulation of the evolution
equations is dependent on a gauge for which $\langle f,v
\rangle=0$. Taking this as well as $d_\mu=\Theta f_\mu +
\nabla_\mu\Theta$ into account, we have
\begin{eqnarray}
&& d_0 \equiv  \langle d, e_0 \rangle = \dnup{\Lambda}{0}{0}\Sigma + \dnup{\Lambda}{0}{a}\underline{e}_a\left(\Omega\right), \\
&& d_a \equiv \langle d, e_a \rangle = \Omega f_a + \dnup{\Lambda}{a}{0}\Sigma + \dnup{\Lambda}{a}{b} \underline{e}_b\left(\Omega\right). 
\end{eqnarray}
Assume that the perturbed hyperboloidal data
$(\mathcal{H},h_{\alpha\beta},K_{\alpha\beta},\Omega,\Sigma)$
satisfies the technical condition
\begin{equation}
\mathring{d}_a - \dnup{\Lambda}{a}{0}\Sigma - \dnup{\Lambda}{a}{b} \underline{e}_b(\Omega) = O(\Omega), \label{technical:condition}
\end{equation}
close to $\partial\mathcal{H}$, where $\mathring{d}_a$ denote the
value of the components of the 1-form $d_\mu$ for the reference
solution and $O(\Omega)$ indicates that the quotient of the left hand side 
divided by $\Omega$ goes to zero as $\Omega$ goes to zero. Initial data for
the congruence of conformal geodesics on which the conformal Gaussian
system hinges will be prescribed such that:
\begin{subequations}
\begin{eqnarray}
&& \dot{x} = e_0,  \label{cg1} \\
&& d_a = \mathring{d}_a, \label{cg2} \\
&& f_a = \Omega^{-1}\left(\mathring{d}_a - \dnup{\Lambda}{a}{0}\Sigma - \dnup{\Lambda}{a}{b} \underline{e}_b(\Omega) \right)\label{cg3}
\end{eqnarray}
\end{subequations}
on $\mathcal{H}$. As a result of the technical condition
(\ref{technical:condition}) one has that that $f_a$ has a finite limit at $\partial \mathcal{H}$
 ---and is thus well defined there. This particular
choice of initial data for the congruence of conformal geodesics provides an easy analysis of the structure of the
conformal boundary of the spacetime ---as will be seen in \ref{section:structure:of:the:conformal:boundary}.
 
Initial values for $\Theta_*$, $\dot{\Theta}_*$, $\ddot{\Theta}_*$ which
characterise the canonical conformal factor $\Theta$ associated to the
congruence of conformal geodesics defined by (\ref{cg1})-(\ref{cg3})
are given on $\mathcal{H}$ by
\begin{equation}
\label{cf:data}
\Theta_*=\Omega,  \quad \dot{\Theta}_* = \langle d,e_0 \rangle,  \quad 2 \Omega \ddot{\Theta}_*= g^\sharp(d,d)_*,
\end{equation}
with the corresponding limits for points on $\partial \mathcal{H}$. As usual,
the subscript ${}_*$ indicates that the value of the functions are
extended of the initial hyperboloid by requiring them to be constant
along a given conformal geodesic with initial data given by
(\ref{cg1})-(\ref{cg3}). It follows from the conformal Hamiltonian
constraint, equation (\ref{conformal_Hamiltonian}) that $d$ is null on
$\partial\mathcal{H}$. Note that as a consequence of the conformal Hamiltonian constraint it follows that when $\Omega=0$
\begin{eqnarray*}
&&\eta^{ij}d_i d_j = \left(\dnup{\Lambda}{0}{0}\right)^2\Sigma^2 - \delta^{ab}\dnup{\Lambda}{a}{c}\underline{e}_c(\Omega) \dnup{\Lambda}{b}{d} \underline{e}_d(\Omega), \\
&& \phantom{\eta^{ab}d_a d_b}= \Sigma^2 + D^\alpha\Omega D_\alpha\Omega =0, 
\end{eqnarray*}
consistently with the choice of initial data for the
conformal factor in (\ref{cf:data}). 

Initial data for the Jacobi field associated to the congruence
of conformal geodesics arising from (\ref{cg1})-(\ref{cg3}) is set by
\[
\eta_{k} =\mathring{\eta}_k =\langle \mathring{\eta}, \mathring{e}_k \rangle, \quad \mbox{ on } \mathcal{H}. 
\]

\bigskip
Finally, one needs to construct tilted data for the rescaled Weyl
($d_{\mu\nu\lambda\rho}$) and Schouten ($L_{\mu\nu}$) tensors. For
this, one starts with the standard hyperboloidal data for these
tensors implied by the conformal constraint equations. As a
consequence of the local existence result given by Proposition \ref{lemma:local existence},
one can regard $\mathcal{H}$ rightfully as a hypersurface of a
spacetime. Let $\underline{d}_{ijkl}$ and $\underline{L}_{ij}$ denote
the components of the Weyl and Schouten tensors with respect to the frame $\{\underline{e}_k\}$. 
The tilted initial data for the Schouten and rescaled Weyl tensors on $\mathcal{H}$ is given by
\[
L_{ij} = \dnup{\Lambda}{i}{n} \dnup{\Lambda}{j}{m} \underline{L}_{nm}, \quad\quad
d_{ijkl} \equiv \dnup{\Lambda}{i}{m} \dnup{\Lambda}{j}{n} \dnup{\Lambda}{k}{p} \dnup{\Lambda}{l}{q} \underline{d}_{mnpq}.
\]

The required spinorial data is then obtained by means of a space
spinor decomposition of the spinorial counterpart of the above tensor.

%

\subsection{Structure of the conformal boundary}
\label{section:structure:of:the:conformal:boundary}

In this subsection we analyse the structure of the conformal boundary
of the development of hyperboloidal data which is close to hyperboloidal
data of the reference radiative spacetime. Attention is focused, in
particular, on the location of timelike infinity $i^+ $. This
investigation is possible as in the vacuum case the conformal factor
can be calculated \textit{a priori} from the initial data.

\medskip
Along the congruence of conformal geodesics with initial data on $\mathcal{H}$ given by (\ref{cg1})-(\ref{cg3}), the canonical conformal factor, $\Theta$,  takes the form
\begin{equation}
\label{quadraticTheta}
\Theta(\tau) = \Omega + \dot{\Theta}_* \tau + \frac{1}{2}  \ddot{\Theta}_* \tau^2,
\end{equation}
with $\dot{\Theta}_*$ and $\ddot{\Theta}_*$ given by the conditions in
(\ref{cf:data}).  As it is customary, denote by $\mathscr{I}$ the
locus of points in the development of the data
$(h_{ij},\chi_{ij},\Theta,\Sigma)$ for which $\Theta=0$. For $q \in
\mathscr{I}$ to qualify as a candidate to be the timelike infinity
of the development it needs to satisfy
\begin{equation} 
\label{i+conditions}
\Theta(q) = 0, \quad \mathrm{d}\Theta(q) = 0, \quad \mbox{Hess} \Theta(q) \mbox{ \;\;non-degenerate.}
\end{equation}
We recall that 
the 1-form $d_\mu = \theta f_\mu + \nabla_\mu \Theta$ and that $2\Theta \ddot{\Theta} = g^\sharp(d,d) $ hold. Thus, for smooth $f_\mu$ and $\Theta$ we observe that on $\mathscr{I}$, $d_\mu$ is null and can only vanish there if $\mathrm{d}\Theta = 0$. Therefore the only point on $\mathscr{I}$ at which all $d_a=\langle d, e_a \rangle$ can vanish, is at a candidate for $i^+$.

\bigskip
Define the function
\[
\Delta \equiv g^\sharp(d_*,d_*) -  \dot{\Theta}_*^2 = -\delta^{ab}d_a d_b,
\] 
and observe that $\Delta = 0$ if and only if $d_a=0$. We recall here that $d_a$, and hence $\Delta$, are constant along conformal geodesics ---see e.g. \cite{Fri95,Fri03c}.

Now, by construction of the reference solution from the initial Cauchy
data given on $\bar{\mathcal{S}}$ and our choice of $\mathcal{B}_a(i)
$, there is a unique point in $\mathcal{B}_a(i) \subset \bar{\mathcal{S}} $ 
such that $\Delta =0$, namely $i$ itself. 
By our choice of coordinates, $i$ lies at the
origin of our spatial coordinate chart and is identified with the
north pole of $\Sphere^3$. Thus by Theorem \ref{Katopart1} and remark
$2$ following it, the point $(\bar{\tau}, x^\mathcal{A})=(0,0,0,0)$ is the
unique point in $D(\mathcal{B}_a(i))$ satisfying the conditions for $i^+ $. This 
corresponds to $(\tau, x^\mathcal{A})=(\bar{\tau}_0,0,0,0)$ in the 
$\mathcal{S}_0$-adapted coordinates. Furthermore the points $(\tau,0,0,0)$ 
are by construction the only points in $D(\mathcal{B}_a(i))$ where $\Delta = 0$ 
for the reference solution. Given our choice for the initial data of 
$d_\mu$ in (\ref{cg2}) for the perturbed spacetime, it follows immediately that the
only point at which $\Delta$ vanishes on $\mathcal{H}$ is the
origin. Therefore the only candidate for $i^+$ is the point, assuming
it exists, where the conformal geodesic $\gamma(\tau) = (\tau,0,0,0) $
intersects $\mathscr{I}$.  Note that for general initial data it is
possible for the development to contain a singularity before reaching
$i^+$ ---as in the case of black hole spacetimes.

\bigskip
The first condition in (\ref{i+conditions}) implies
\begin{equation}
\label{scri_tau}
\tau = \frac{-\dot{\Theta}_* \pm \sqrt{-\Delta}}{\ddot{\Theta}_*},
\end{equation}
while the second one can be rewritten as
\[
\mathrm{d}\Omega + \mathrm{d}\dot{\Theta}_* \tau + \frac{1}{2} \mathrm{d}\ddot{\Theta}_* \tau^2 + (\dot{\Theta}_* + \ddot{\Theta}_* \tau)\mathrm{d}\tau = 0.
\]
Together with $\Delta = 0$, both conditions imply that 
\begin{equation}
\label{dTheta1}
\tau =\tau_+ \equiv - \frac{\dot{\Theta}_* }{ \ddot{\Theta}_*} = -\frac{\Omega}{\dot{\Theta}_*}.
\end{equation}
For sufficiently small perturbations, $\dot{\Theta}_* = \langle d, v
\rangle = \dnup{\Lambda}{0}{0}\Sigma +
\dnup{\Lambda}{0}{a}\underline{e}_a\left(\Omega\right)$ does not
vanish and $\tau_+$ is finite. In fact later on the initial data will
be restricted such that $\tau_+ $ lies in the interval $(0,2
\bar{T}_0)$. Note that for the reference solution we have $\tau_+ =
\bar{T}_0$ and $q=i \in \bar{\mathcal{S}}$. By construction we have
that $\mathrm{Hess}\Theta(q) $ is non-degenerate, and hence the
$i^+$-candidate in the reference spacetimes can be regarded,
rightfully, as future timelike infinity.

\subsubsection{Behaviour of the Hessian of $\Theta$} 
The point $q=(\tau_+,0,0,0)$ satisfies the first two conditions of
(\ref{i+conditions}).  We now verify that the conditions discussed in
the previous paragraphs imply that the Hessian of $\Theta$ is
non-degenerate at $q$. For this we make use of the following
expression:
\begin{eqnarray*}
&& \nabla_\mu \nabla_\nu \Theta =
 \nabla_\mu \nabla_\nu  \Omega + \nabla_\mu \nabla_\nu  \dot{\Theta}_* \tau + \frac{1}{2} \nabla_\mu \nabla_\nu \ddot{\Theta}_* \tau^2 \nonumber  \\
 && \hspace{2cm} + 2 \nabla_{(\mu} \dot{\Theta}_* \nabla_{\nu)} \tau + 2 \tau \nabla_{(\mu} \ddot{\Theta}_* \nabla_{\nu)} \tau  \nonumber \\
&& \hspace{2cm}  + (\dot{\Theta}_* + \ddot{\Theta}_*\tau) \nabla_\mu \nabla_\nu  \tau + \ddot{\Theta}_* \nabla_\mu\tau  \nabla_\nu  \tau. 
\end{eqnarray*}
Using the conformal Gaussian coordinates $(\tau, x^\mathcal{A})$ for our calculations and $\alpha = \langle b,v \rangle_* = \dot{\Theta}_*/\Omega$ as well as (\ref{dTheta1}) we find that
\begin{subequations}
\begin{eqnarray}
\label{Hess1}
&& \nabla_0 \nabla_0 \Theta\vert_{i^+} 
=\frac{1}{2}\Omega \alpha^2,
\\
\label{Hess2}
&& \nabla_0 \nabla_{\mathcal{A}} \Theta\vert_{i^+} = -\Omega \nabla_{\mathcal{A}} \alpha,
\\
\label{Hess3}
&& \nabla_{\mathcal{A}} \nabla_{\mathcal{B}} \Theta\vert_{i^+} =
 \frac{2}{\Omega\alpha^2}\eta^{ab} \nabla_{\mathcal{A}} d_a \nabla_{\mathcal{B}} d_b 
+  \frac{2\Omega}{\alpha^2} \nabla_{\mathcal{A}}\alpha \nabla_{\mathcal{B}}\alpha,
\end{eqnarray}
\end{subequations}
where the right hand side is given in terms of initial data and
spatial derivatives. Observe that $\nabla_{\mathcal{A}} d_a$ are
identical to the ones in the reference solution and thus the only
change arises from the perturbation of $\alpha$. For data
$\dot{\Theta}_*, \Omega $ close enough to the reference solution, in
the sense that at the spatial origin $\alpha$ and $\nabla_{\mathcal{A}}\alpha $ are
sufficiently close to their counterparts in the reference solution,
the Hessian will not be degenerate and thus $q$ can be identified as
the timelike infinity $i^+ $ of the perturbed spacetime. Note that one
may rescale the initial data such that $\Sigma$ remains fixed, while
$\Omega = \mathring{\Omega}$. This greatly simplifies the analysis of
the effect of perturbations of $\alpha$ on the Hessian.

\subsection{The stability result for purely radiative spacetimes}
\label{section:stability:result}
We are now in the position of presenting our main result ---a
stability result for purely radiative spacetimes arising from
hyperbolidal data. The proof of this result follows the ideas of
\cite{Fri88,LueVal09}.

\bigskip
Let $\mathring{u}_0$ and $u_0$ denote, respectively, the tilted initial
data for the radiative reference spacetime and for a perturbation
thereof, as discussed in the previous section. Using the
diffeomorphism $\mathring{\phi}$, we can think of $\mathring{u}_0$ and
$u_0$ as a vector-valued function over $\Sphere^3$. On the other hand,
using $\phi$, $u_0$ can only be regarded as a
vector valued function over $\mathcal{U}\subset \Sphere^3$. On
$\mathcal{U}$, let
\[
w \equiv u_0 - \mathring{u}_0.
\]
Using the linear extension operator given in section \ref{section
extending initial data}, with $Ew $ satisfying (\ref{norm:relationship}), we set on $\Sphere^3$
\[
\breve{u}_0\equiv Ew, \quad u_0 = \mathring{u}_0 + \breve{u}_0.
\]
Consistently with the previous discussion of the initial data, one writes
\begin{equation}
\label{Ansatz:perturbation}
u = \mathring{u}+ \breve{u}, 
\end{equation}
where we recall that both $\mathring{u}$ and $\breve{u}$ are regarded as vector
valued functions over $\Real \times \Sphere^3$. In particular, the quantities in
\[
\breve{u} \equiv (\breve{e}^\es_{AB}, \breve{f}_{AB}, \breve{\xi}_{ABCD}, \breve{\chi}_{(AB)CD}, \breve{\Theta}_{ABCD}, \breve{\phi}_{ABCD}, \breve{\eta}, \breve{\eta}_{AB}),
\]
describe the (non-linear) perturbations of the reference radiative
spacetime. Substituting the Ansatz (\ref{Ansatz:perturbation}) into
the evolution system (\ref{sh:system}) one obtains a symmetric
hyperbolic system for $\breve{u}$ of the form
\begin{equation}
\label{sh:system:perturbation}
A^0(\mathring{u}+ \breve{u}) \cdot \partial_\tau \breve{u} + \sum_{\er=1}^3 A^\er(\mathring{u}+\breve{u}) \cdot c_\er (\breve{u}) + \hat{B}(\tau,x^\mathcal{A},\mathring{u},\breve{u})\cdot \breve{u}=0, 
\end{equation}
with $\hat{B}(\tau,x^\mathcal{A},\mathring{u},\breve{u})$ a matrix
valued function with entries which are polynomials in $\breve{u}$ of
at most degree one and coefficients which are smooth functions on
$\Real \times \Sphere^3$. As discussed earlier, the matrix valued
functions $A^\es(z)$ are symmetric, have entries which are polynomial
in $z$ of at most degree one and have constant coefficients.  In
contrast to the situation discussed in \cite{LueVal09}, the matrix
$A^0(\mathring{u}_0)$ is not diagonal. Nevertheless, Theorem
\ref{Katopart1} guarantees that there exists a neighbourhood
$D_{\bar{T}_0}\subset D^m_\delta$ such that $\mathring{u}_0 \in
D_{\bar{T}_0}$. Hence,
\[
(z,A^0(\mathring{u}_0)z)>\delta(z,z), \quad \forall z\in\Real^N.
\] 

\bigskip
The previous discussion allows us to formulate an existence and
stability result for the solutions to equation
(\ref{sh:system:perturbation}).

\begin{theorem}
\label{Katopart2}
Suppose $m\geq 4$. Let $u_0=\mathring{u}_0 + \breve{u}_0$ be
hyperboloidal initial data satisfying the technical condition
(\ref{technical:condition}). Given $T in (\bar{\tau}_0, 2\bar{\tau}_0)$, there exists
$\varepsilon>0$ such that:
\begin{itemize}

\item[(i)] For $|| \breve{u}_0||_m < \varepsilon$ there exist a
 unique solution $u=\mathring{u} + \breve{u}$ to the conformal propagation
equations (\ref{p1})-(\ref{p7}) and (\ref{extended_cfe4}) with minimal
existence interval $\tau\in[0,T]$ and $u\in C^{m-2}([0,T]\times
\Sphere^3)$.

\item[(ii)] The associated congruence of conformal geodesics contains
no conjugate points in $[0,T]$.

\item[(iii)] At the origin one has $\tau_+ = -\Omega/\dot{\Theta}_*\in [0,T]$ .

\item[(iv)] The Hessian as given by (\ref{Hess1})-(\ref{Hess3}) is nondegenerate at $(\tau_+,0,0,0) $.

\end{itemize}

\noindent
The solution $u=\mathring{u}+\breve{u}$ on
$\mathcal{D}^+(\mathcal{S})$ implies a $C^{m-2}$ solution
$(\mathcal{M}, \tilde{g}) $ to the vacuum Einstein field equations
with vanishing cosmological constant, where $\tilde{g}_{\mu\nu} = \Theta^{-2}g_{\mu\nu}$ with $\Theta$ given by (\ref{quadraticTheta}).
\end{theorem}

\medskip
\noindent
\textbf{Proof.}  By hypothesis, we know that $\mathring{u}_0 \in
D_{\bar{T}_)} \subset D^m_\delta$. We thus satisfy the conditions of
the variation of Kato's theorem given in \cite{Fri88,LueVal09}. Parts
(i) and (ii) follow directly from Kato's theorem. Parts (iii) and (iv)
follow from the discussion in section
\ref{section:structure:of:the:conformal:boundary}. Due to Lemma
\ref{lemma:propagation constraints} on the propagation of the
extended conformal constraints, one
knows that the existence of a solution of the propagation
system (\ref{p1})-(\ref{p7}) and (\ref{extended_cfe4}) implies a
solution to the full extended conformal Einstein field equations
which in turns implies a
solution to the vacuum Einstein field equations with vanishing
cosmological constant. \hfill $\Box$

\bigskip
One has the following corollary.
\begin{corollary}
\label{Katopart2a}
Suppose the condition of Theorem \ref{Katopart2} are satisfied, then a
solution $(\mathcal{M}, \tilde{g}) $, as given above, has a conformal
boundary given by the set $\Theta=0 $. The conformal boundary consists
of the set $\mathscr{I}$, which represents future null infinity, and
the point $i^+$ given by $(\tau_+,0,0,0) $, which represents timelike
infinity.
\end{corollary}

Hence the conformal boundary of radiative spacetimes is shown to be
stable, subject to the conditions of Theorem \ref{Katopart2}.

\begin{figure}[t]
\centering
 \includegraphics[width=.6\textwidth]{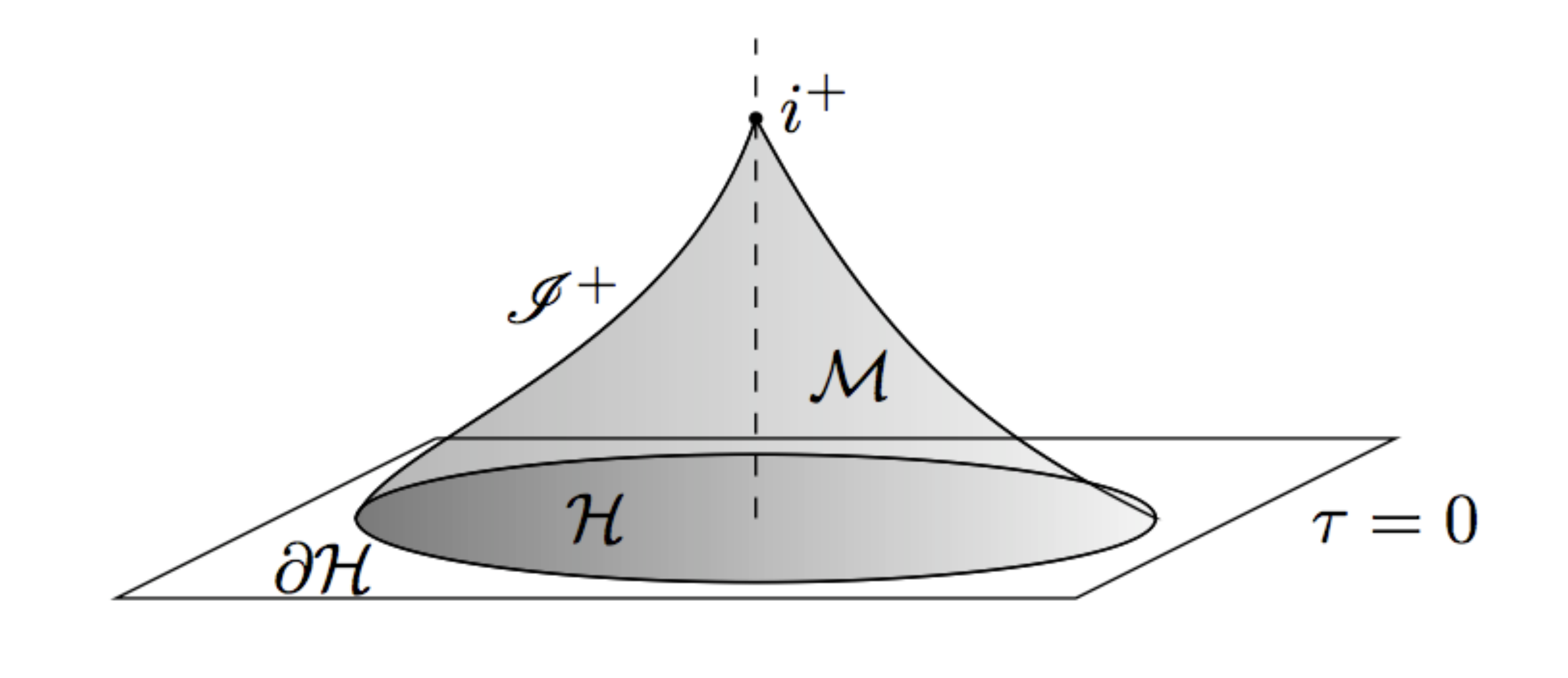}
\caption{The perturbed radiative spacetime and its conformal boundary. Note that this figure is not a conformal diagram.}
\end{figure}


\bigskip

\noindent \textbf{Remark.} Note that the evolution of $u_0$ implies a
unique solution $u$ and a spacetime, which is locally isometric with
$(\mathcal{N}, g_{\mu\nu})$ as obtained in Proposition
\ref{lemma:local existence}.

\section{Conclusions}
The conformal Einstein field equations have been used to formulate an
hyperboloidal initial value problem by means of which one can address
the nonlinear stability of a class of purely radiative
spacetimes. Our analysis is particularly concerned with the
structure of null infinity and the location of future timelike
infinity, $i^+$. The use of a conformal Gaussian gauge system
coordinates allows to calculate the location of the conformal boundary
directly from the initial data, as well as analyse its structure. The
results are a generalisation of previous results on Minkowski-like
spacetimes \cite{LueVal09}. In particular the conditions on the
location of $i^+$ are given in a more general form, while the choice 
of initial data for the perturbed conformal geodesic congruence 
facilitates the stability analysis of the conformal boundary in 
comparison to \cite{LueVal09}.

The reference purely radiative spacetimes are constructed from static
initial data using procedure introduced by Friedrich
\cite{Fri88}. Recent results on the necessary and sufficient
convergence conditions required on a series of \emph{multipolar
data}, one can conclude that there is an infinite
family of reference purely radiative spacetimes. 

We observe that we were able to prove the stability of radiative 
spacetimes and study their conformal boundary without requiring 
any explicit knowledge of the reference spacetime. All our analysis 
has been carried out in the abstract only relying on general  
properties of these spacetimes. This raises the hope that similar 
results may be possible for other classes of spacetimes.

\section{Acknowledgements}
CL would like to thank the Leverhulme Trust for a research project
grant (F/07 476/AI).  JAVK is an EPSRC Advanced Research Fellow.  We
would like to thank Thomas B\"ackdahl for helpful discussions on
multipole moments.


\end{document}